\documentclass[sigconf]{acmart}

\usepackage{booktabs} 

\setcopyright{rightsretained}

\acmDOI{10.475/123_4}

\acmISBN{123-4567-24-567/08/06}

\usepackage{todonotes}
\usepackage{ifpdf}

\usepackage{graphicx}

\usepackage{color}
\usepackage{amsmath}
\usepackage{ amssymb }
\usepackage{amsfonts}
\usepackage{mathrsfs}
\usepackage{dsfont}
\usepackage{amsbsy}
\usepackage{stmaryrd}

\usepackage{algorithm}
\usepackage{algpseudocode}

\usepackage{xpatch}
\usepackage[tight,footnotesize]{subfigure}
\usepackage{url}
\usepackage{tikz}
\usetikzlibrary{arrows,positioning,automata,circuits.logic.US,circuits.logic.IEC}
\newcommand{\coord}{ {\color{\green} coord } }
\newcommand{\router}{ {\color{\rouge} router} } 
\newcommand{\device}{ {\color{\blu} end\_dev } }

\definecolor{lavander}{cmyk}{0,0.48,0,0}
\definecolor{violet}{cmyk}{0.79,0.88,0,0}
\definecolor{burntorange}{cmyk}{0,0.52,1,0}

\def\blu{blue!45!black!70!}
\def\green{green!45!black!70!}

\def\rouge{red!50!black!70!}
\newcommand{\wh}[1]{\color{white}{#1}}
\xpatchcmd{\algorithmic}{\setcounter}{\algorithmicfont\setcounter}{}{}
\providecommand{\algorithmicfont}{}

\newcommand{\ev}[1]{\mathrm{F}_{#1}}
\newcommand{\glob}[1]{\mathrm{G}_{#1}}
\newcommand{\until}[1]{\mathrm{U}_{#1}}
\newcommand{\somewhere}[2]{\Diamonddot_{#1}^{#2}}
\newcommand{\everywhere}[2]{\boxbox_{#1}^{#2}}
\newcommand{\surround}[2]{\circledcirc_{#1}^{#2}}
\newcommand{\reach}[2]{\mathcal{R}_{#1}^{#2}}
\newcommand{\escape}[2]{\mathcal{E}_{#1}^{#2}}

\newcommand{\since}[1]{\mathrm{S}_{#1}}
\newcommand{\fmon}{\mathbf{m}}

\begin{document}
\title{Monitoring Mobile and Spatially \\ Distributed Cyber-Physical Systems}

\author{Ezio Bartocci}
\affiliation{%
  \institution{Technische Universit\"at Wien}
  \city{Wien} 
  \state{Austria} 
}

\author{Luca Bortolussi}
\affiliation{%
  \institution{Universit\`a di Trieste}
  \city{Trieste} 
  \state{Italy} 
}

\author{Michele Loreti}
\affiliation{%
  \institution{Universit\`a di Firenze }
  \city{Florence} 
  \state{Italy} 
}

\author{Laura Nenzi}
\affiliation{%
  \institution{Technische Universit\"at Wien}
  \city{Wien} 
  \state{Austria} 
}

\renewcommand{\shortauthors}{E. Bartocci et al.}

\begin{abstract}
Cyber-Physical Systems~(CPS) consist of collaborative, 
networked and tightly intertwined computational (logical) 
and physical components, each operating at different spatial 
and temporal scales.
Hence, the spatial and temporal 
requirements play an essential role for their correct and safe execution. 
Furthermore, the local interactions among the system components result 
in global spatio-temporal emergent behaviors often 
impossible to predict at the design time.  
In this work, we pursue a complementary approach by introducing STREL
a \emph{novel spatio-temporal logic} that enables the specification of
spatio-temporal requirements and their monitoring 
over the execution of \emph{mobile} and \emph{spatially distributed} CPS.  
Our logic extends the Signal Temporal Logic~\cite{MalerN13}  with two novel spatial 
operators \emph{reach} and \emph{escape} from which is possible to  
derive other spatial modalities such as \emph{everywhere}, \emph{somewhere}
 and \emph{surround}. These 
 operators enable a monitoring procedure where 
 the satisfaction of the property at each location depends 
 only on the satisfaction of its neighbours, opening the way 
 to future distributed online monitoring algorithms. We propose  
 both a \emph{qualitative} and \emph{quantitative} semantics based on 
 \emph{constraint semirings}, an algebraic structure suitable for 
 constraint satisfaction and optimisation. We prove that, for 
 a subclass of models, all the spatial properties expressed with 
 \emph{reach} and \emph{escape}, using euclidean distance, 
 satisfy all the model transformations using rotation, reflection and 
 translation. Finally, we provide an offline monitoring algorithm 
 for  STREL and, to demonstrate  the feasibility of our approach,
 we show its application using the monitoring of a simulated
 mobile ad-hoc sensor network as running example.
\end{abstract}

%
%




\maketitle

\let\thefootnote\relax\footnotetext{This is the authors' version of the accepted manuscript published in the 
Proceedings of the 15th ACM-IEEE International Conference on Formal Methods and Models for System Design
Vienna, Austria — September 29 - October 02, 2017, available at the following DOI: \href{https://doi.org/10.1145/3127041.3127050}{https://doi.org/10.1145/3127041.3127050}}

\section{Introduction}

%

\label{sec:intro}
From micro- and nano-scale cyber and physical/biological materials to self-driving cars, smart factories and smart cities, cyber-physical systems (CPS) are reshaping the way 
in which we perceive and interact with our physical world, 
becoming ubiquitous in our society. CPS consist of collaborative, 
networked, spatially distributed, and tightly intertwined computational 
(logical) and physical components, each operating at different spatial 
and temporal scales.  Therefore, the spatial and the temporal requirements 
are fundamentals for their safe and correct execution.

The  openness of CPS with the possibility for new actors to join or to 
leave the system, the local interactions among the system components 
and the unknown environment in which they operate may cause undesired 
spatio-temporal emergent behaviours (i.e., congestion) often impossible 
to predict at the design-time. Indeed, their complexity restricts the 
exhaustive verification of their models runtime only to relatively small examples. 
Here, we pursue a complementary approach by introducing the 
\emph{Spatio-Temporal Reach and Escape Logic} ({\textsc STREL}), a novel 
formal specification language that enables to express in a concise 
way complex spatio-temporal requirements and to monitor them for the 
first time (to the best of our knowledge) over the execution of 
\emph{mobile and spatially  distributed CPS.} 
 
The idea of the proposed framework stems from the attempt to generalise 
and to overcome some limitations of the \emph{Spatio-Signal Temporal Logic} (SSTL) 
previously introduced in~\cite{NenziBCLM15}.  SSTL extends the Signal 
Temporal Logic~\cite{MalerN13} with modalities (named \emph{somewhere} 
and \emph{surround}) to express also \emph{spatial properties} and
 it is interpreted over a discrete model of the space, represented 
as a finite undirected graph.  Each node represents a location in 
the space, characterised by a set of signals whose evolution can be observed 
in time, while each edge of the graph is labelled with a positive weight, 
that can be used to represent the distance between two nodes.  This provides 
a metric structure to the space in terms of shortest path distances, 
enabling to monitor also spatial properties.  However, since the topology 
of the graph in SSTL is assumed to be static, one main limitation is 
the impossibility to monitor nodes changing locations.
Furthermore, monitoring of spatial properties is performed on each location by changing the graph so to consider only the locations that satisfy the distance constraint. This means that, 
the monitoring results of a location cannot be \emph{reused} in
the monitoring of its neighbours. In this work, we decide to completely reformulate the spatial 
modalities changing the perspective: instead of searching locations 
satisfying properties within a certain distance using the shortest path, 
the satisfaction of a location can be obtained by \emph{using} monitored 
values obtained from the directly connected locations.

In particular, STREL generalizes SSTL by considering two 
new operators, named \emph{reach} and \emph{escape}.  These new operators 
simplify the monitoring procedure that can be computed 
\emph{locally}: the satisfaction of the property at each location 
depends only on the satisfaction of its neighbours\footnote{We will see 
in Section~\ref{sec:alg} that this feature is very important to define distributed and online monitor algorithms.}.  
Furthermore, while SSTL operates on spatio-temporal models 
that are static (the locations do not change their positions), 
{\textsc STREL} can handle also mobile/dynamic CPS.  We also 
prove that, for a subclass of models, all the spatial properties 
expressed with \emph{reach} and \emph{escape}, using euclidean 
distance, satisfy all the transformed models through rotation, reflection and 
 translation.
 
Another important feature of our logic considered in this paper is that, 
following an approach similar to the one considered in~\cite{LM05}, 
we do not rely on a specific domain for interpreting logical properties. 
Indeed, STL/SSTL semantics can be either \emph{qualitative}, 
ranging over \emph{boolean values}, or \emph{quantitative}, 
ranging over \emph{real values}.
In this paper, we propose both qualitative and quantitative 
semantics based on \emph{Constraint Semirings}.
These are algebraic structures that consist of a domain and 
two operations named \emph{choose} and \emph{combine}. 
Constraint semirings have been shown to be very flexible, expressive and 
convenient for a wide range of problems, in particular for optimization and 
solving problems with soft constraints and multiple criteria~\cite{BMR97}. 
The use of semirings allows the definition of a single monitoring 
procedure that, being parametric with respect to the class of 
data collected from devices and values produced as results, 
can be used with different purposes.  
We then provide an offline monitoring algorithm for STREL, and, to
illustrate the main features of the proposed formal framework,  
we show its application using the monitoring of a simulated
{\it Mobile Ad-hoc sensor NETwork} (MANET) as our running example.

We want to stress that  STREL is a flexible framework to formulate properties of CPS:  the ability of freely mixing spatial and temporal operators to build complex queries, and to automatically construct monitoring algorithms, marks a neat difference from other related approaches, like the development of ad hoc solutions for specific properties.

The rest of this paper is organized as follows. 
Section~\ref{sec:related}  discusses the related work.  
Section~\ref{sec:def} introduces the model we 
consider to represent the spatio-temporal signals, while 
section~\ref{sec:ReachSTL} provides the syntax and the semantics of STREL.  
An offline monitoring algorithm and its implementation is then discussed in section~\ref{sec:alg}.  
In section~\ref{sec:results}, we show the logic at work on some examples, in particular we consider a MANET as case study. Section~\ref{sec:conclusion} draws our conclusions 
and discusses future works.

\section{Related Work}
\label{sec:related}

Monitoring spatial-temporal properties over CPS executions
was first proposed in~\cite{Talcott08} where the author 
has introduced the notion of spatial-temporal event-based 
model for CPS.  Events are triggered 
by the execution of actions, by the exchange of messages and 
by physical changes. Each generated event is labeled 
with time and space stamps and processed 
by a monitor. 
In~\cite{TVG09}, this concept is further elaborated, developing 
 a spatial-temporal event-based model where the space is 
 represented as a 2D Cartesian coordinate system with location 
points and location fields. 

The approaches described in~\cite{Talcott08,TVG09} provide an 
algorithmic framework enabling a user to develop manually a 
monitor. However, they do not provide any spatio-temporal logic 
language enabling the specification and the automatic 
monitoring generation.

In the field of \emph{collective adaptive systems}~\cite{CianciaLLM16}, 
other mathematical  structures, such as \emph{topological spaces}, \emph{closure spaces}, 
\emph{quasi-discrete closure spaces} and \emph{finite graphs}~\cite{NenziBCLM15}, 
have been considered to reason about spatial relations, such as \emph{closeness} 
and \emph{neighborhood}.  
Despite these models are suitable for offline and centralised monitoring of model-based simulations, 
they do not scale well for the runtime monitoring of spatially distributed CPS.

Several logic-based formalisms have been proposed to specify the 
behavior and the spatial structure of concurrent systems~\cite{CC04} and 
for reasoning about the topological~\cite{BC02} or directional~\cite{BS10} 
aspects of the interacting entities. 
 In topological reasoning~\cite{BC02}, the spatial objects are sets of points 
 and the relation between them is preserved under translation, scaling 
 and rotation. 
 In directional reasoning, the relation between objects depends on 
 their relative position. These logics are usually highly computationally 
 complex~\cite{BS10} or even undecidable~\cite{MR99}. 
 
Monitoring spatial-temporal behaviors has started to receive more attention 
only recently with SpaTeL~\cite{bartocci2015} and SSTL~\cite{NenziBCLM15}. 
The {\it Spatial-Temporal Logic} (SpaTeL)~\cite{bartocci2015} is the unification of 
{\it Signal Temporal Logic}~\cite{MalerN13} (STL) and {\it Tree Spatial Superposition Logic} (TSSL)
introduced in~\cite{bartocci2014,Bartocci2016} to classify and detect spatial patterns. TSSL reasons over quad trees, spatial data
structures that are constructed by recursively partitioning the space into uniform quadrants. The notion of
superposition in TSSL provides a way to describe statistically the distribution of discrete states in a particular
partition of the space and the spatial operators corresponding to \emph{zooming in and out in} a particular region of
the space. By nesting these operators, it is possible to specify self-similar and fractal-like structures \cite{GrosuSCWEB09} that generally characterize the patterns emerging in nature.  
The procedure allows one to capture very complex spatial structures, but at the price of a complex formulation of spatial properties, which are in practice only learned from some template image.

Another important work to mention is \textsc{Voltron}~\cite{MottolaMWG14}, an open-source  \emph{team-level programming} system 
for drone's collaborative sensing.  \textsc{Voltron} provides special programming constructs to reason about time and space
and allows users to express sophisticated collaborative tasks without exposing them to the complexity 
of concurrent programming, parallel execution, scaling, and failure recovery.  
The spatial constructs are limited to operate on a set of locations of a given geometry (that the user needs to specify).  
The system is suitable more for programming than for monitoring. For example, it does not allow 
to quantify how much the current CPS execution  is close to violate a given requirement.

\newcommand{\wfun}{\mathbf{W}}
\newcommand{\nextto}[3]{#1\stackrel{#2}{\mapsto}#3}
\newcommand{\route}{\tau}
\newcommand{\tsign}{\nu}
\newcommand{\pct}{\tilde{\tsign}}
\newcommand{\ssign}{\mathbf{s}}
\newcommand{\sts}{\sigma}
\newcommand{\pcsts}{\tilde{\sigma}}
\newcommand{\lserv}{\lambda}

\section{Spatial Models, Signals and Traces}
\label{sec:def}

In this section, we introduce the model of space we consider, 
and the type of signals that the logic specifies.

\subsection{Constraint Semirings}

An elegant and general way to represent the result of monitoring is based 
on \emph{constraint semiring}. This is an algebraic structure that consists 
of a domain and two operations named \emph{choose} and \emph{combine}. 
Constraint semirings are subclass of semirings which have been shown 
to be very flexible, expressive and convenient for a wide range of problems, 
in particular for optimisation and solving problems with soft constraints 
and multiple criteria~\cite{BMR97}, and in model checking~{\protect{\cite{LM05}}}.

\begin{definition}[semiring]
A \emph{constraint semiring} (just \emph{semiring} in the following) 
is a tuple $\langle A, \oplus, \otimes, \bot, \top \rangle$ composed by 
a set $A$, two operators $\oplus$, $\otimes$ and two constants $\bot$, 
$\top$ such that: 
\begin{itemize}
\item $\oplus : 2^A \rightarrow A$ is an associative, commutative, 
idempotent operator to ``choose'' among values\footnote{We let 
$x\oplus y$ to denote $\oplus(\{ x , y\})$.}, with $\oplus(\emptyset)=\top$;
\item $\otimes : A \times A \rightarrow A$ is an associative, commutative 
operator  to ``combine'' values;
\item $\otimes$ distributes over $\oplus$;
\item $\bot \oplus a = a$, $\top \oplus a = \top$, $\top \otimes a = a$, 
$\bot \otimes a = \bot$ for all $a \in A$;
\item $\sqsubseteq$, which is defined as $a\sqsubseteq b$ iff {$a\oplus b=b$}, 
provides a complete lattice $\langle A , \sqsubseteq , \bot, \top \rangle$. 
\end{itemize}
We say that a \emph{semiring} $A$ is \emph{idempotent} if and only if for 
any $a\in A$ $a\oplus a=a\otimes a =a$. Moreover, we say that a \emph{semiring}
$A$ is \emph{total} when $\sqsubseteq$ is a {\emph{total order}}.

\end{definition}


With an abuse of notation we sometimes refer to a semiring 
$\langle A, \oplus,\otimes , \bot, \top  \rangle$ with the carrier $A$ 
and to its components by subscripting them with the carrier, i.e., 
$\oplus_A$, $\otimes_A$, $\bot_A$ and $\top_A$.  For the sake of a lighter 
notation we drop the subscripts if clear from the context.


\begin{example}\label{ex:semirings}
Typical examples of semirings that we will use in this paper are\footnote{We use $\mathbb{R}^{\infty}$ (resp. $\mathbb{N}^{\infty}$) to denote $\mathbb{R}\cup\{-\infty,+\infty\}$ (resp. $\mathbb{N}\cup\{\infty\}$).}:
\begin{itemize}
%
\item the Boolean semiring $\langle  \{\mathit{true},\mathit{false}\}, \vee, \wedge, \mathit{false}, \mathit{true} \rangle$; 
\item the tropical semiring $\langle \mathbb{R}_{\geq 0}^{\infty},\emph{min},+,+\infty,0 \rangle$;
\item the max/min semiring: $\langle \mathbb{R}^{\infty}, \emph{max},\emph{min}, -\infty, +\infty \rangle$ ;
\item the integer semiring: $\langle \mathbb{N}^{\infty}, \emph{max},\emph{min}, 0, +\infty \rangle$.
%
%
\end{itemize}
Boolean, max/min and integer semirings are \emph{idempotent} while tropical semiring is not. All the above semirings are \emph{total}. 

\end{example}

One of the advantages of \emph{semirings} is that these can be easily composed. For instance, if $A$ and $B$ are two semirings, one can consider the \emph{cartesian product} $\langle A\times B,(\bot_A,\bot_B), (\top_A,\top_B), \oplus,\otimes\rangle$ where operations are applied elementwise.

\subsection{Spatial model}

Space is represented via a graph with edges having a weight from a given semiring.

\begin{definition}
    Let $\langle A, \oplus, \otimes, \bot, \top \rangle$ be a \emph{semiring}, a  $A-$\emph{spatial model} $\mathcal{S}$ is a pair $\langle L, \wfun\rangle$ where:
    \begin{itemize}
        \item $L$ is a set of \emph{locations}, also named \emph{space universe};
        \item $\wfun\subseteq L\times A\times L$ is a \emph{proximity function} associating at most one label $w \in A$ with each distinct pair $\ell_1,\ell_2\in L$. 
    \end{itemize} 
    
\end{definition}
We will use $\mathbb{S}_{A}$ to denote the set of $A$-\emph{spatial models}, while $\mathbb{S}^{L}_{A}$ indicates the set of $A$-\emph{spatial models} having $L$ as a set of locations. In the following, we will equivalently write  $(\ell_1,w,\ell_2)\in \wfun$ as $\wfun(\ell_1,\ell_2)=w$ or  $\nextto{\ell_1}{w}{\ell_2}$, saying that $\ell_1$ is \emph{next to} $\ell_2$ with weight $w \in A$.

A special class of spatial models are the ones based on \emph{Euclidean spaces}. 

\begin{definition}[Euclidean spatial model] 
\label{def:Euclidean}
Let $L$ be a set of locations, $R\subseteq L\times L$ a (reflexive) relation and $\mu: L\rightarrow \mathbb{R}^{2}$ a function mapping each location to a point in  $\mathbb{R}^{2}$, we let $\mathcal{E}(L,R,\mu)$ be the $\mathbb{R}^{\infty}\times\mathbb{R}^{\infty}$-spatial model\footnote{$\mathbb{R}^{\infty}$ is the \emph{min/max} semiring considered in Example~\ref{ex:semirings}.} $\langle L, \wfun^{\mu, R}\rangle$ such that:
\[
\wfun^{\mu,R}=\{ (\ell_1,\mu(\ell_1)-\mu(\ell_2),\ell_2) | (\ell_1,\ell_2)\in R \}
\]
\label{def:euclisomod}
\end{definition}
\vspace{-5mm}
Note that we label edges with a 2-dimensional vector $w$ describing how to reach $\ell_2$ from $\ell_1$, i.e.,  $\mu(\ell_1) + w = \mu(\ell_2)$. This obviously allows us to compute the euclidean distance between  $\ell_1$ and $\ell_2$ as $\| w \|_2$, but, as we will see, allows us to compute the euclidean distance of any pair of locations connected by any path, not necessarily by a line in the plane.
%
%
%
%
%
%
%
%
%

\begin{example}[Mobile Ad hoc sensor NETwork]
\label{ex:manet} 
A Mobile Ad-hoc sensor NETwork (MANET) is a sensor network that can consist of  up ten thousands of mobile devices connected wirelessly. The devices are usually deployed to monitor environmental changes  such as pollution, humidity, light and temperature.
Each sensor node can be equipped with a sensing transducer, data processor, a radio transceiver  and an embedded battery.  It can move independently in any direction and change its links to other devices. 
Two nodes can communicate each other if  their Euclidean distance is at most their  communication range as depicted in Fig.~\ref{fig:proxconnect}~(right) .
Moreover, the nodes can be of different type and their behaviour and communication can depend on their types.
\\
When considering a MANET, we can easily define different proximity functions for the same set of locations, where each location represents a mobile device.
Given a set of $n$ reference points in a two-dimensional Euclidean plane, a Voronoi 
diagram~\cite{Aurenhammer1991}  partitions the plane into set of $n$ regions, one per reference point,  
assigning each point of the plane to the region corresponding to the closest reference point. 
The dual of the  Voronoi diagram is the proximity graph or Delaunay triangulation~\cite{Delaunay1934}.  
In Figure~\ref{fig:proxconnect} (left), we can see an example of Voronoi diagram (in  blue) and proximity graph (in red). 
The proximity function can then be defined with respect to the Cartesian coordinates, as in Definition~\ref{def:euclisomod}: $\wfun^{\mu,R}(\ell_i,\ell_j)=\mu(\ell_i)-\mu(\ell_j)=(x_i,y_i)-(x_j,y_j)= (x_i - x_j , y_i -y_j)$, where $(x_i,y_i)$ are the plane coordinates of the location $\ell_i$.
\\
The proximity function can be also equal to a value that depends of other specific characteristics or behaviours of our nodes. For instance, Fig.~\ref{fig:proxconnect}~(right) represents the connectivity graph of MANET. In this case a location $\ell_i$ is next to a location $\ell_j$ if and only if they are within their communication range.

\begin{figure}[H]
    \centering
    \includegraphics[scale=0.33]{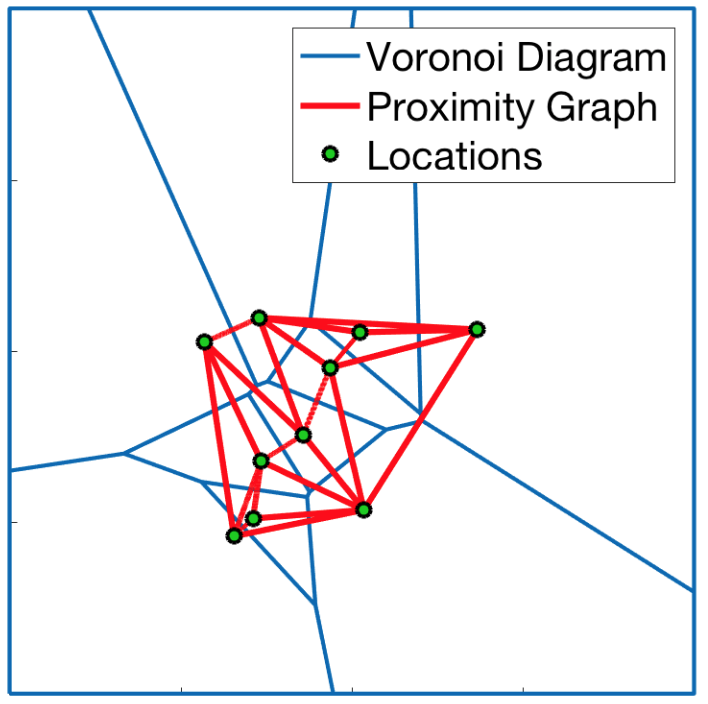}
    \includegraphics[scale=0.33]{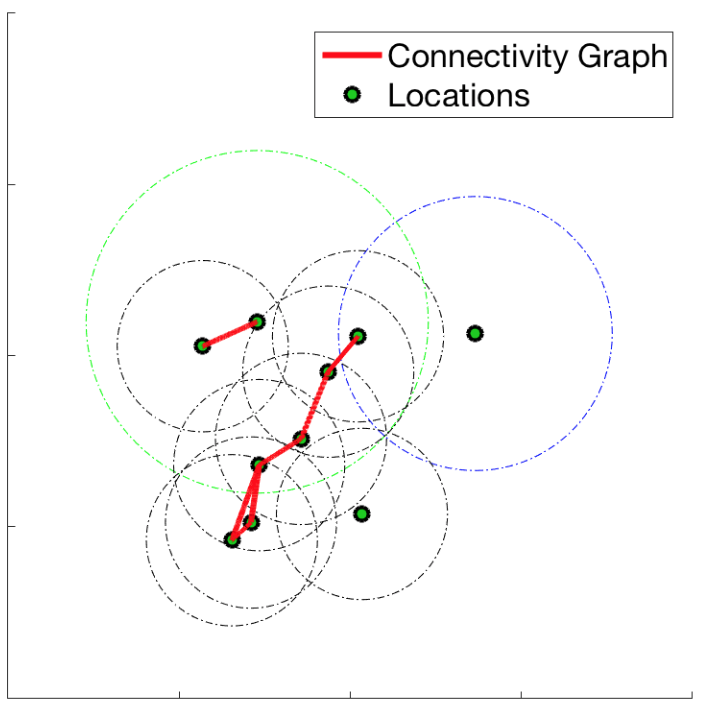}
    \caption{{\small Proximity graph (left) and Connectivity graph (right)}}
    \label{fig:proxconnect}
\end{figure}

\end{example}

Given an $A$-spatial model we can define \emph{routes}. 

\begin{definition}
Let $\mathcal{S}=\langle L,\wfun\rangle$, a \emph{route} $\route$ 
is an infinite sequence $\ell_0 \ell_1\cdots \ell_k\cdots$ in $L^{\omega}$ such that for any $i\geq 0$, $\nextto{\ell_i}{d}{\ell_{i+1}}$.
\end{definition}
Let $\route=\ell_0 \ell_1\cdots \ell_k\cdots$ be a route, $i\in \mathbb{N}$ and $\ell\in L$, we use:
\begin{itemize}
\item $\route[i]$ to denote the $i$-th node $\ell_i$ in $\route$;
\item $\route[i..]$ to indicate the suffix route $\ell_i \ell_{i+1} \cdots$;
\item $\ell \in \route$ when there exists an index $i$ such that $\route[i]=\ell$, while we use $\ell\not\in \route$ if this index does not exist;
\item $\route(\ell)$ to denote the first occurrence of $\ell$ in $\rho$:
\[
\route(\ell)=\left\{
\begin{array}{ll}
\min\{ i | \route[i]=\ell \} & \mbox{if $\ell\in \route$}\\
\infty & \mbox{otherwise} \\ 	
\end{array}
\right.
\]
\end{itemize}
We also use $Routes(\mathcal{S})$ to denote the set of routes in $\mathcal{S}$, while $Routes(\mathcal{S},\ell)$ denotes the set of routes starting from $\ell \in L$.

We can use routes to define the \emph{distance} among two locations in a \emph{spatial model}. This distance is computed via an appropriate function $f$ that combines all the weights in a route into a value taken from an appropriate semiring $B$. 

\begin{definition}
	Let $\mathcal{S}=\langle L,\wfun\rangle$ be an $A$-spatial model, $\route$ a route in $\mathcal{S}$, $\langle B,\oplus_{B},\otimes_{B},\bot_{B},\top_{B}\rangle$ a \emph{complete} semiring and $f:B\times A\rightarrow B$ a {\it distance monotone function} such that $b \sqsubseteq_{B} f(b,a)$, or $f(b,a) \sqsubseteq_{B} b$, for any $a\in A$ and $b\in B$. The distance $d_{\route}^{f}[i]$ up-to index $i$ is:
	\[
	d_{\route}^{f}[i]= \begin{cases}
      \bot_{B} &   i=0 \\
      f(d_{\route[1..]}^{f}[i-1],w) & (i>0)\mbox{ and } \nextto{\route[0]}{w}{\route[1]} 
\end{cases} \\	
	\]
	
\noindent	
Given a locations $\ell\in L$, the distance over $\route$ up-to $\ell$ is then $d_{\route}^{f}(\ell)  = d_{\route}^{f}[\route(\ell)]$ if $\ell\in \route$, or $\top_{B}$ otherwise.
%
\end{definition}


\begin{example}
\label{ex:distancefunction}
Considering again a MANET, one could be interested in different types of distances, e.g., 
\emph{counting} the number of \emph{hops},  or distances induced by the weights of the Euclidean space structure.
%

\noindent
To count the number of hops, we can simply use the  function $hops: \mathbb{N}^{\infty}\times \mathbb{R}^{\infty}_{\geq 0}\rightarrow \mathbb{N}^{\infty}$, taking values in the tropical semiring on $\mathbb{N}^{\infty}$:
\[
hops(v,w)=v+1
\]   
and in this case $d^{hops}_\tau[i]=i$.

Considering the proximity function $\wfun^{\mu,R}(\ell_i,\ell_j)$ computed from the Cartesian coordinates, we can use the distance induced by the function $\Delta$ defined as follow	
		\[
		\Delta(v,(x,y))= v + \| (x,y) \|_2,
		\]
where $(x,y)$ are the  coordinates of the vectors returned by $\wfun^{\mu,R}$ while $v$ is the distance incrementally computed by $\Delta$. It is easy to see that for any route $\route$  and for any  location $\ell \in L$ in $\route$, the function $d_{\route}^{\Delta}(\ell)$ yields the sum of lengths of the edges in $\mathbb{R}^{2}$ connecting $\ell $ to $\route(0)$.

Both the functions $hops$ and $\Delta$ are \emph{monotone} and satisfy the constraints:
\[
  hops(v,w) \sqsubseteq_{\mathbb{N}} v \qquad 
\Delta(v,(x,y))  \sqsubseteq_{\mathbb{R}^{\infty}_{\geq 0}}  v 
\]
 
%
%

The distance between two locations $\ell_1$ and $\ell_2$ is obtained by choosing the distance values along all possible routes starting from $\ell_1$ and ending in $\ell_2$, according to the $\oplus$ operation of the semiring $B$:
\[d_{\mathcal{S}}(\ell_1,\ell_2) = \oplus\{ d_{\route}(\ell_2) | \route\in Routes(\mathcal{S},\ell_2) \}.
\]
\begin{example}
\label{ex:distancefunction}
Consider again the distance functions defined for a MANETS. For \emph{hops}, we are taking the minimum hop-length over all paths connecting $\ell_1$ and $\ell_2$, resulting in the shortest path distance. In the Euclidean case, the function $\Delta$ returns the same result along any path, which will also be our distance, due to idempotence of $\oplus$.
\end{example}

	
%
%
%
%
\end{example}	

\subsection{Spatio-Temporal Signals}

\begin{definition}
A 
{\emph{signal domain}} is a tuple $\langle D, \oplus,\otimes, \odot,\top,\bot\rangle$ where:
\begin{itemize}
	\item $\langle D, \oplus,\otimes, \top,\bot\rangle$, is an \emph{idempotent semiring};
	\item $\odot: D\rightarrow D$, is a \emph{negation function} such that:
	\begin{itemize}
	\item $\odot\top =\bot$;
	\item $\odot\bot = \top$;
	\item $\odot(v_1\oplus v_2)=(\odot v_1)\otimes (\odot v_2)$
	\item $\odot(v_1\otimes v_2)=(\odot v_1)\oplus (\odot v_2)$
	\item for any $v\in D$, $\odot ( \odot v ) = v$.
	\end{itemize}	
\end{itemize}	
\end{definition}

In this paper, we will consider two \emph{signal domains}:
\begin{itemize}
	\item Boolean signal domain $\langle \{ \top , \bot \}, \vee, \wedge,\neg \rangle$ for qualitative monitoring;
	\item {Max/min signal domain $\langle \mathbb{R}^{\infty}, \max, \min, -\rangle$} for quantitative monitoring.
\end{itemize}

For signal domains we will use the same notation and notational conventions introduced for semirings. 

\begin{definition} Let $\mathbb{T}=[0, T]$ a time domain and $\langle D, \oplus,\otimes, \odot,\top,\bot\rangle$ a \emph{signal domain}, a \emph{temporal $D$-signal} $\tsign$ is a function
$\tsign: \mathbb{T}\rightarrow D$.

\noindent
Consider a finite sequence: 
\[
\pct = [(t_{0}, d_0),\ldots,(t_{n}, d_{n})]
\]
such that, for any $i \in \{0,\ldots,n\}$, $t_i<t_{i+1}$ and $d_i\in D$. Usually, $t_{0}=0$.
We let $\pct$ denote a \emph{piecewise constant temporal $D$-signal} in $\mathbb{T}=[0, T]$, that is 
  \[
  \pct(t) = \begin{cases}
      & \bot   \quad \text{ for } t < t_{0}, \\
      & d_i  \quad \text{ for } t_{i} \leq t < t_{i+1}, \\
      & d_n \quad \text{ for } t_{n} \leq T;
\end{cases} \\
\]
\end{definition}

Given a \emph{piecewise constant temporal signal} $\pct=[(t_{0}, d_0),\ldots,(t_{n}, d_{n})]$ we will use $\mathcal{T}( \pct )$ to denote the set $\{ t_0,\ldots, t_n \}$ of \emph{time steps} in $\pct$; $start(\pct)$ to denote $t_0$; while we will say that $\pct$ is \emph{minimal} if and only if for any $i$, $d_i\not=d_{i+1}$. 
We will also let $\pct[ t=d ]$ to denote the signal obtained from $\pct$ by adding the element $(t,d)$. 
Finally, if $\tsign_1$ and $\tsign_2$ are two $D$-temporal signals, and $op: D\times D\rightarrow D$, $\tsign_1~op~\tsign_2$ denotes the signal associating with each time $t$ the value $\tsign_1(t)~op~\tsign_2(t)$. Similarly, if $op:D_1 \rightarrow D_2$, $op~\tsign_1$ denotes the $D_2-$signal associating with $t$ the value $op~ \tsign_1(t)$.

\begin{definition} Let $L$ be a \emph{space universe}, and $\langle D, \oplus,\otimes, \odot,\top,\bot\rangle$ a signal domain. A \emph{spatial $D$-signal} is a function $\ssign: L\rightarrow D$.  
\end{definition}

\begin{definition}[Spatio-temporal $D$-signal]
  Let $L$ be a space universe, $\mathbb{T}=[0, T]$ a time domain, and $\langle D, \oplus,\otimes, \odot,\top,\bot\rangle$ a signal domain,   a spatio-temporal $D$-signal is a function
  \[ \sts: L \rightarrow \mathbb{T} \rightarrow D \]
\noindent
such that $\sts(\ell)=\tsign$ is a temporal signal that returns a value $\tsign(t) \in {D}$ for each time $t  \in \mathbb{T}$. We say that $\sts$ is \emph{piecewise constant} when for any $\ell$, $\sts(\ell)$ is a \emph{piecewise constant temporal signal}. \emph{Piecewise constants spatio-temporal} signal are denoted by $\pcsts$. 
\end{definition}

Given a spatio-temporal signal $\sts$, we will use $\sts@t$
to denote the \emph{spatial signal} at time $t$, i.e. the signal $\ssign$ such that $\ssign(\ell)=\sts(\ell)(t)$, for any $\ell \in L$.
Different kinds of signals can be considered while the signal domain $D$ is changed. Signals with  $D= \{ true , false \}$  are called \emph{boolean signals}; with $D = \mathbb{R}^{\infty}$ are called real-valued or \emph{quantitative signals}.

 \begin{definition}[$D$-Trace]
Let $L$ be a space universe, a {\it spatio-temporal $D$-trace} is a function
$$\vec x: L \rightarrow \mathbb{T} \rightarrow D^{n}$$ 
such that for any $\ell\in L$ yields a vector of  temporal signals $\vec{x}(\ell)=(\tsign_1,\ldots,\tsign_n)$.
In the rest of the paper we will use $\vec{x}(\ell,t)$ to denote $\vec{x}(\ell)(t)$.
\end{definition}

We plan to work with spatial models that can dynamically change their configurations. For this reason, we need to define a function that returns the spatial configuration at each time.

\begin{definition}[Location service]
Let $L$ be a spatial universe, a \emph{location service} is a function $\lserv: \mathbb{T}\rightarrow \mathbb{S}^L_{A}$ associating  each element in the time domain $\mathbb{T}$ with a spatial model $\mathbb{S}^L_{A}$ that describes the spatial configuration of locations. 
\end{definition}

\begin{example}
Let us considering a MANET with a proximity graph. A $\mathbb{R}^{\infty}-$spatio temporal signal $\sts: L \rightarrow \mathbb{T} \rightarrow \mathbb{R}^{\infty}$ associates a temporal signal $\sts(i)=\tsign$ of real-values at each location $\ell \in L=\{\ell_1,\dots,\ell_7\}$; $\sts@t$ instead corresponds to the spatial signal at time $t$, i.e. it is a function that returns a value $\sts(\ell_i)(t)$ for each location $\ell_i$ at time t.
We can see the use of the location service in the figure below.
The plot shows two different spatial configurations of the model for time $t_1$ and $t_2$. We can see that locations $\ell_1$ and $\ell_2$ change their position, this changes also the Voronoi diagram and the proximity graph. We have then two different proximity functions on the same space universe $L$, i.e. $\lserv(t_1)=\langle L,\wfun_1 \rangle$, $\lserv(t_2)=\langle L,\wfun_2 \rangle$.
\begin{figure}[H]
    \centering
    \includegraphics[height=4.15cm]{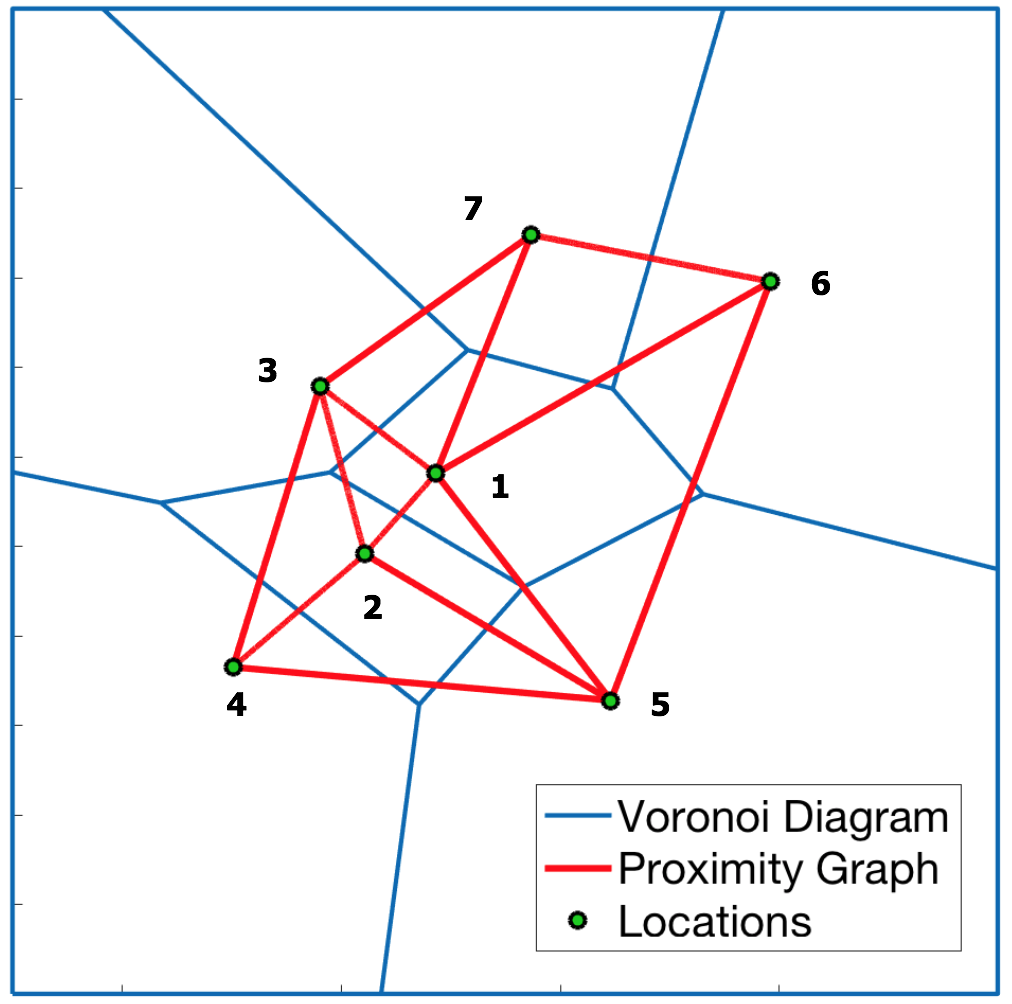}
    \includegraphics[height=4.15cm]{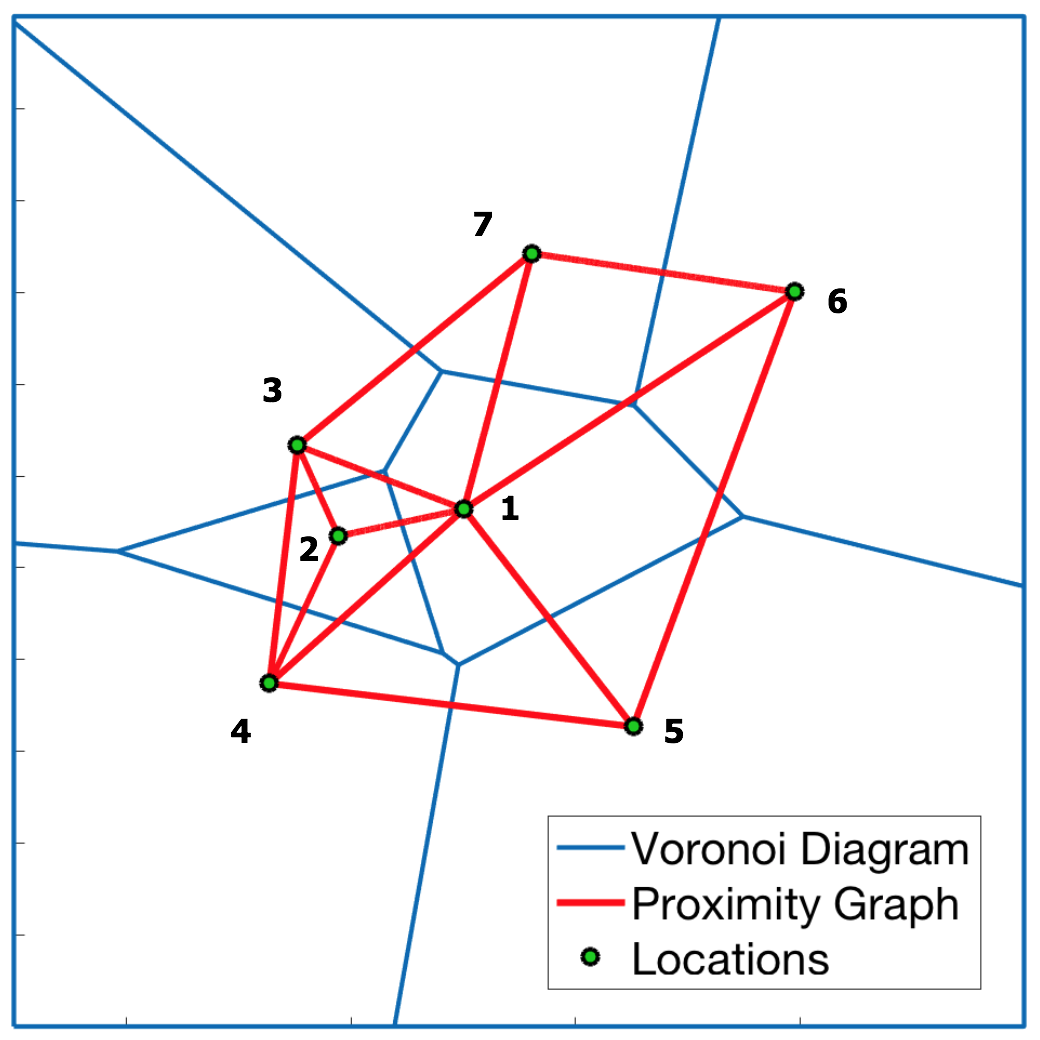}
    \caption{Two snapshots of a spatial model with 7 locations $\ell_1,\dots,\ell_7$ that move in a 2D Euclidian space. The plane is partitioned using a Voronoi Diagram ({\color{blue} blue}). In {\color{red} red} we have the proximity graph.}
    \label{fig:voronoimobility}
\end{figure}
\end{example}

\section{Spatio-temporal Reach and Escape Logic}
\label{sec:ReachSTL}
In this section,  we present the {\it Spatio-Temporal Reach 
and Escape Logic}~(STREL), an extension of the {\it Signal Temporal Logic}. 
We define the syntax and the semantics of STREL, describing 
in detail the spatial operators and their expressiveness. 

\subsection{Syntax}

The syntax of STREL is given by
{\small
\[
\varphi :=  
    \mu \mid  
    \neg \varphi \mid  
    \varphi_{1} \wedge \varphi_{2} \mid  
    \varphi_{1} \: \until{[t_{1},t_{2}]} \: \varphi_{2} \mid 
    \varphi_{1} \: \since{[t_{1},t_{2}]} \: \varphi_{2} \mid  
    \varphi_{1} \: \reach{ d}{f} \: \varphi_{2} \mid 
    \escape{d}{f}  \: \varphi  
\]
}
where  $\mu$ is an {\it atomic predicate} ($AP$), {\it negation} $\neg$ and {\it conjunction} $\wedge$ are the standard Boolean connectives, $\until{[t_{1},t_{2}]}$ and  $\since{[t_{1},t_{2}]}$ are the {\it Until} and the {\it Since} temporal modalities, with $[t_{1},t_{2}]$ a real positive closed interval. These are the standard temporal operators of STL, and we refer  the reader to~\cite{MalerN13, Donze2013} for more details.
The spatial modalities are the   {\it reachability} $\reach{d}{f}$ and the {\it escape} $\escape{ d}f{}$ operators, with $f$ a \emph{Distance Function}, (we call $DF$ their collection), described in the previous section, and $d$ a \emph{Distance Predicate} (from a set $DP$ of predicates), e.g., inequalities\footnote{With an abuse of notation, we will denote by $\neg d$ the predicate that complements $d$.}. The exact meaning of $f$ and $d$ depends on specific interpretation functions. This because, the monitored value associated with a formula $\varphi$ depends on the considered domain. We impose that any $d$ occurring in a operator $\reach{d}{f}$ is $\sqsubseteq-closed$, i.e., if $x$ satisfies $d$ and $y\sqsubseteq x$ then $y$ satisfies $d$. This because predicate $d$ represents an \emph{upper bound} on a distance. 

The reachability operator $\phi_1 \reach{d}{f}\phi_2$ describes the behavior of reaching a location satisfying property $\phi_2$ passing only through locations that satisfy  $\phi_1$, through nodes whose distance from the initial location satisfy the predicate $d$.
The escape operator $\escape{d}f{\phi}$, instead, describes the possibility of escaping from a certain region passing only through locations that satisfy $\phi$, via a route with distance satisfying the predicate $d$. Differently from $\reach{}{}{}$, in $\escape{d}{f}$ the predicate $d$ represents a \emph{lower bound}.  For this reason we assume that any interpretation of $d$ is $\sqsupseteq-closed$, i.e., if $x$ satisfies $d$ and $x\sqsubseteq y$ then $y$ satisfies $d$\footnote{$d$ is $\sqsubseteq-closed$ if and only if $\neg d$ is $\sqsupseteq-closed$.}.

As customary, we can derive the {\it disjunction} operator $\vee$ and the future  {\it eventually} $\ev{[t_{1},t_{2}]}$ and  {\it always} $\glob{[t_{1},t_{2}]}$ operators from  the until temporal modality, and the corresponding past variants from the since temporal modality, see~\cite{MalerN13} for details. 
We can define also other three derived spatial operators: the {\it somewhere} and the {\it everywhere} that describe behaviors of some or of all locations at a certain distance from a specific point, and the {\it surround} that expresses the topological notion of being surrounded by a $\phi_2$-region, while being in a $\phi_1$-region, with additional metric constraints. A more thorough discussion of the spatial operators will be given after introducing the semantics.

\subsection{Semantics}
The semantics of STREL is  evaluated point-wise  at each time and each location. We stress that each STREL formula $\varphi$ abstracts from the specific domain used to express the satisfaction value of $\varphi$ as well as there is not explicit reference to the semiring used in the spatial model to express \emph{weights} associated with edges. These, of course, are needed to define the semantics.  In the following, we assume that $D_1$ is the domain of the spatio-temporal traces, while $D_2$ is the semiring where the logic is evaluated. Furthermore, $A$ is the semiring of weights, and $B$ is the semiring in which distance functions take values. To define the semantics, we also need three auxiliary functions.
The \emph{signal interpretation function} $\iota: AP\times D_1^{n} \rightarrow D_2$  permits to translate the input trace in a different  ${D}_2$-spatio temporal signal, for each atomic proposition in $AP$, which will be the input of the monitoring procedure.
The function $\gamma: DF\rightarrow (B\times A \rightarrow B)$ is used to interpret function symbols as proper distance functions, while  $\delta: DB\rightarrow (B\rightarrow \{ true, false \})$ maps distance predicate symbols into proper predicates.

\begin{definition} [Semantics] 
\label{generalsemantics}
Let $A$ and B be two semirings, and $D_1$ and $D_2$ two signal domains.
Let $L$ be a \emph{space universe}, $\vec x$ be a {\it spatio-temporal $D_1$-trace} for $L$ and
$\lserv: \mathbb{T} \rightarrow \mathbb{S}^{A}_{L}$  the \emph{location service} associating an A-spatial model $\mathbb{S}^L_{A}$ at each time in $\mathbb{T}.$
Let $\iota$, $\gamma$, and $\delta$ be the functions introduced above. 
The $D_2$-monitoring function $\fmon$ of $\vec x$ is recursively defined in Table~\ref{tab:monitoring}.

\end{definition}

\begin{table*}
\begin{center}	
\begin{tabular}{rcl}
$\fmon( \lserv, \vec{x}, \mu, t, \ell)$ & $=$ & $\iota(\mu,\vec{x}(t,\ell))$ \\[.2cm]

$\fmon(\lserv, \vec{x}, \neg\varphi, t, \ell)$ & $=$ & $\odot_{D_{2}} \fmon(\lserv, \vec{x}, \varphi, t, \ell)$  \\[.2cm]

$\fmon( \lserv, \vec{x}, \varphi_1 \wedge \varphi_2, t, \ell)$ & $=$ & $\fmon( \lserv, \vec{x}, \varphi_1, t, \ell) \otimes_{D_2} \fmon(\lserv, \vec{x}, \varphi_2, t, \ell)$  \\[.2cm]  

$\fmon( \lserv, \vec{x}, \varphi_{1} \: \until{[t_{1},t_{2}]} \: \varphi_{2}, t, \ell)$ & $=$ & ${\bigoplus_{D_2}}_{t' \in [t+t_{1}, t+t_{2}]} \big (\fmon(  \lserv, \vec{x}, \varphi_2, t', \ell) \otimes_{D_2} {\bigotimes_{D_2}}_{t'' \in [t, t']} \fmon( \lserv, \vec{x}, \varphi_1, t'', \ell) \big)   $ \\[.2cm]
      
$\fmon( \lserv, \vec{x}, \varphi_{1} \: \since{[t_{1},t_{2}]} \: \varphi_{2}, t, \ell)$ & $=$ &  
${\bigoplus_{D_2}}_{t' \in  [t-t_{2}, t-t_{1}]} \big (\fmon(\lserv, \vec{x}, \varphi_2, t', \ell) \otimes_{D_2} {\bigotimes_{D_2}}_{t'' \in [t', t]} \fmon(  \lserv, \vec{x}, \varphi_1, t'', \ell) \big)   $ \\[.2cm]

 $\fmon(\lserv, \vec{x}, \varphi_{1} \: \reach{d}{f} \: \varphi_{2}, t, \ell)$ & $=$ & 
 ${\bigoplus_{D_2}}_{\tau\in Routes(\lserv(t),\ell)}
        ~~{\bigoplus_{D_2}}_{\ell'\in\tau :\delta(d)\left(d_{\tau}^{\gamma(f)}(\ell')\right)}
        \left(
            \fmon(  \lserv, \vec{x}, \varphi_2, t, \ell')
            \otimes_{D_{2}}
            {\bigotimes_{D_2}}_{j < \tau(\ell')} 
                \fmon(  \lserv, \vec{x}, \varphi_1, t, \tau[j])             
        \right)$ \\[.2cm]

$\fmon( \lserv, \vec{x}, \escape{d}{f} \: \varphi, t, \ell)$ & $=$ &
${\bigoplus_{D_2}}_{\tau\in Routes(\lserv(t),\ell)}
        ~~{\bigoplus_{D_2}}_{\ell'\in \tau:\delta(d)\left(d_{\lambda(t)}^{\gamma(f)}(\ell,\ell')\right)}
            ~~{\bigotimes_{D_2}}_{i \leq \tau(\ell')} 
                \fmon(  \lserv, \vec{x}, \varphi, t, \tau[i])$ \\[.2cm]
\end{tabular}
\end{center}

	\caption{Monitoring function.}
	\label{tab:monitoring}
\end{table*}

Given a formula $\phi$, the function $\fmon( \lserv, \vec{x}, \phi, t, \ell)$ corresponds to the evaluation of the formula at time $t$ in the location $\ell$. 
The choice of $B, D_2, \iota, \gamma$ and $\delta$ produces different types of semantics. As described in Section ~\ref{sec:def}, we consider two signal domains: $\mathbb{B}$ and $\mathbb{R^\infty}$, giving rise to qualitative and quantitative monitoring, correspond respectively to a Boolean answer value and real satisfaction value. 
We describe the semantics for the 
Boolean signal domain ($D_2 = \langle \{ \top , \bot \}, \vee, \wedge,\neg \rangle $ ). 
We say that $(\lambda , x(\ell,t))$ satisfies a formula $\phi$ if $\fmon( \lserv, \vec{x}, \phi, t, \ell)= \top$. 
The procedure will be exactly the same for different choices of the formula evaluation domain, just operators have to be interpreted according to the chosen semirings and signal domains.
We use the following example as the system on which we specify our properties, in particular we will use  the graph in Figure~\ref{fig:spprop} to describe the spatial operators.

\begin{example}[ZigBee protocol]
\label{ex:zigbee}
 In Fig.~\ref{fig:spprop}, the graph represents a MANET. In particular, we consider the nodes with three different  roles such as the ones implemented in the ZigBee protocol: {\it coordinator}, {\it router} and {\it EndDevice}. The Coordinator node $(\coord)$, represented in green color in the graph, is unique in each network and is responsible to initialize the network.  After the initialisation, the coordinator behaves as a router. 
The Router node $(\router)$, represented in red color in the graph, acts as a intermediate router, passing on data from other devices. The EndDevice node $(\device)$, represented in blue, can communicate 
only with a parent node (either the Coordinator or a Router) and it is unable to relay data from other devices.
Nodes move in  space and the figure corresponds to the spatial configuration at a fixed time $t$.  As trace and location service, let us consider a $\mathbb{R}^{\infty}$-spatial model as the proximity graph presented in  Example \ref{ex:manet} and a  $\mathbb{B}$-trace over this graph $\vec x: L \rightarrow \mathbb{T} \rightarrow \mathbb{B}^{3}$ denoting the kind of node, i.e. $\vec x (\ell, t)=(\top,\bot,\bot)$ if $\ell$ is a coordinator, $\vec x (\ell, t)=(\bot,\top,\bot)$ if $\ell$ is a router, and  $\vec x (\ell, t)=(\bot,\bot,\top)$ if $\ell$ is an end node. 
\end{example}

\noindent{\bf Atomic Proposition.} $\fmon( \lserv, \vec{x}, \mu, t, \ell)=\iota(\mu,\vec{x}(t,\ell)).$ Different types of atomic propositions and signal interpretations are admissible. We can simply consider a finite set 
$\{p_1, \dots, p_n \}=AP$ and an interpretation function $\iota(p_i,\vec x(\ell,t))=\top$ iff $x_i(\ell,t)=\top$. E.g., in Fig.~\ref{fig:spprop}, we can consider atomic propositions describing the type of node, i.e., the boolean propositions $\{ \coord, \router, \device \}$ are true if the node is of the corresponding type. In case of real valued signals and of a quantitative interpretation of the logic ($D_2$ being in this case the real valued max/min semiring), we  can consider inequalities $\mu=(g(\vec{x})\geq 0)$ for some real function $g$ and define $\iota(\mu,\vec{t,\ell})=g(\vec{x,t})$.

\noindent{\bf Negation.} $\fmon(\lserv, \vec{x}, \neg\varphi, t, \ell)= \neg \fmon( \lserv, \vec{x}, \varphi, t, \ell)$ 

\noindent{\bf Conjunction.} {\small $\fmon( \lserv, \vec{x}, \varphi_1 \wedge \varphi_2, t, \ell)$ $=$ $\fmon( \lserv, \vec{x}, \varphi_1, t, \ell)$ $\wedge$ $\fmon(\lserv, \vec{x}, \varphi_2, t, \ell)$}

\noindent{\bf Until.} $\fmon( \lserv, \vec{x}, \varphi_{1} \until{[t_{1},t_{2}]}\varphi_{2}, t, \ell) = \bigvee_{t' \in t + [t_{1}, t_{2}]}  (\fmon(  \lserv, \vec{x}, \varphi_2, t', \ell) \wedge \bigwedge_{t'' \in [t, t']} \fmon( \lserv, \vec{x}, \varphi_1, t'', \ell) \big) $.
As customary, $(\lambda , x(\ell,t))$ satisfies \\ $\varphi_{1} \until{[t_{1},t_{2}]} \varphi_{2}$ iff it satisfies $\varphi_{1}$ from $t$ until, in a time between $t_{1}$ and $t_{2}$ time units in the future, $\varphi_{2}$ becomes true. Note how the temporal operators are evaluated in each location separately.

\noindent{\bf Since.} $\fmon( \lserv, \vec{x}, \varphi_{1} \: \since{[t_{1},t_{2}]} \: \varphi_{2}, t, \ell)  = \bigvee_{t' \in t - [-t_{2}, -t_{1}]} \linebreak \big (\fmon(\lserv, \vec{x}, \varphi_2, t', \ell) \wedge \bigwedge_{t'' \in [t', t]} \fmon(  \lserv, \vec{x}, \varphi_1, t'', \ell) \big)$.
 $(\lambda , x(\ell,t)$ satisfies $\varphi_{1} \: \since{[t_{1},t_{2}]} \: \varphi_{2}$ iff it satisfies $\varphi_{1}$ from now since, in a time between $t_{1}$ and $t_{2}$ time units in the past, $\varphi_{2}$ was true.

Except for the interpretation function, the semantics of the boolean and the temporal operators is directly derived from and coincident with that of STL (qualitative for Boolean signal domain and quantitative for an $\mathbb{R}^\infty$ signal domain), see~\cite{Donze2013} for details. 


\noindent{\bf Reachability.} {\small $\fmon(\lserv, \vec{x}, \varphi_{1} \: \reach{d}{f} \: \varphi_{2}, t, \ell)=
\bigvee_{\tau\in Routes(\lserv(t),\ell)} \linebreak
        \bigvee_{\ell'\in \tau:\delta(d)\left(d_{\tau}^{\gamma(f)}(\ell')\right)} 
        ( \fmon(  \lserv, \vec{x}, \varphi_2, t, \tau(\ell'))
            \wedge 
            \bigwedge_{j < \tau(\ell')}  \fmon(  \lserv, \vec{x}, \varphi_1, t, \tau[j]) )$}

\noindent $(\lambda , x(\ell,t))$ satisfies
$\varphi_{1} \: \reach{d}{f} \: \varphi_{2}$ iff it satisfies $\varphi_2$ in a location $\ell'$ reachable from $\ell$ through a route $\tau$, with a length $d_{\tau}^{\gamma(f)}(\ell')$ satisfying the predicate $\delta(d)$, and such that $\tau[0]=\ell$ and all its elements with index less than $\tau(\ell')$ satisfy $\varphi_1$. 
 In Figure~\ref{fig:spprop}, we report an example of reachability property, considering f as the $hops$ function described in Example~\ref{ex:distancefunction}. In the graph, the location  $\ell_6$ (meaning the trajectory $\vec{x}$ at time t in position $\ell_6$) satisfies $\device \: \reach{m \leq 1}{hops} \: \router$, with distance predicate $d = m \leq 1$ being true if the distance is less than or equal to 1 units. Indeed, there exists a route $\tau = \ell_6\ell_5$ such that $d_{\tau}^{hops}[1]=1$, where $\tau[0]=\ell_6$, $\tau[1]=\ell_5$, $\tau[1]$ satisfies the red property (it is a router) and all the other elements of the route satisfy the blue property (they are end-devices). Instead, for example, the location $\ell_8$ does not satisfy the property because it does not satisfies the blue (end-device) property.
 
\noindent{\bf Escape.} $\fmon( \lserv, \vec{x}, \escape{d}{f} \: \varphi, t, \ell)= 
\bigvee_{\tau\in Routes(\lserv(t),\ell)} \linebreak
        \bigvee_{\ell'\in \tau:\delta(d)\left(d_{\lambda(t)}^{\gamma(f)}(\ell,\ell')\right) }
            ~\bigwedge_{i \leq \tau(\ell')} 
                \fmon(  \lserv, \vec{x}, \varphi, t, \tau[i]).
$
\noindent $(\lambda , x(\ell,t))$
              satisfies  $\escape{d}{f} \: \varphi$ if and only if there exists a route $\route$ and a location $\ell'\in \route$ such that $\route[0]=\ell$ and $d_\mathcal{S}(\tau[0],\ell')$ satisfies the predicate $\delta(d)$, while $\ell'$ and all the elements $\tau[0],...\tau[k-1]$ (with $\route(\ell')=k$) satisfy $\varphi$.             
In Fig~\ref{fig:spprop}, we report an example of escape property. In the graph,  the location $\ell_{10}$ satisfies $ \escape{m \geq 2}{hops} \: \neg \device$. Indeed, there exists a route $\tau = \ell_{10}\ell_7\ell_8$ such that $\tau[0]=\ell_{10}$, $\tau[2]=\ell_8$, $d_S^{hops}(\ell_{10},\ell_1)=2$ and $\ell_{10}$, $\ell_7$ and $\ell_8$ do not satisfy the blue property, i.e. they are not end-devices. Note that the route $\ell_{10}\ell_{11}\ell_{16}$ is not a good route to satisfy the property because the distance $d_S^{hops}(\ell_{10},\ell_{16})=1$. 

\begin{figure}[h]
{\small
\center
\begin{tikzpicture}
  [scale=.6,auto=left,every node/.style={circle,thick,inner sep=0pt,minimum size=6mm}]
  \node (1) [fill=\blu, draw = black] at (-3,-3) {\wh{1}};
  \node (2) [fill=\blu, draw = black] at ( 1,-2) {\wh{2}};
  \node (3) [fill=\blu, draw = black] at ( 3,-1) {\wh{3}};
  \node (4) [fill=\blu, draw = black] at ( -3, 1) {\wh{4}};
  \node (5) [fill=\rouge, draw = black] at ( 1, 2) {\wh{5}};
  \node (6) [fill=\blu, draw = black] at (-1, 2) {\wh{6}};
  \node (7) [fill=\rouge, draw = black] at (0, 0) {\wh{7}};
  \node (8) [fill=\rouge, draw = black] at (-2,-1) {\wh{8}};
  \node (9) [fill=\rouge, draw = black] at (3,3) {\wh{9}};
  \node (10) [fill=\green, draw = black] at (4,1) {\wh{10}};
  \node (11) [fill=\rouge, draw = black] at (5,0) {\wh{11}};
  \node (12) [fill=\blu, draw = black] at (6,-2) {\wh{12}};
  \node (13) [fill=\blu, draw = black] at (8,1) {\wh{13}};
  \node (14) [fill=\blu, draw = black] at (5,3) {\wh{14}};
  \node (15) [fill=\blu, draw = black] at (8,-1) {\wh{15}};
  \node (16) [fill=\rouge, draw = black] at (6.5,1.8) {\wh{16}};

 \draw [-] (1) -- (8) node[midway] {};
 \draw [-] (2) -- (7) node[midway] {};
 \draw [-] (8) -- (6) node[midway] {};
  \draw [-] (8) -- (7) node[midway] {};
 \draw [-] (7) -- (10) node[midway] {};
 \draw [-] (7) -- (5) node[midway] {};
 \draw [-] (3) -- (10) node[midway] {};
  \draw [-] (6) -- (5) node[midway] {};
 \draw [-] (10) -- (11) node[midway] {};
 \draw [-] (10) -- (9) node[midway] {};
 \draw [-] (11) -- (15) node[midway] {};
 \draw [-] (11) -- (12) node[midway] {};
\draw [-] (9) -- (14) node[midway] {};
\draw [-] (10) -- (14) node[midway] {};
\draw [-] (10) -- (16) node[midway] {};
\draw [-] (11) -- (16) node[midway] {};
\draw [-] (13) -- (16) node[midway] {};
\draw [-] (8) -- (4) node[midway] {};

 \end{tikzpicture}
\caption{
Example of spatial properties. {\bf Reachability:} $\device\: \reach{m\leq 1}{hops} \: \router$. {\bf Escape:} $ \escape{m\geq 2}{hops} \: \neg \device$.
{\bf Somewhere}: $\somewhere{m\leq 4 }{hops} \coord$.  {\bf Everywhere}:    $\everywhere{m\leq 2 }{hops} \router$.   {\bf Surround:} $ (\coord \vee \router ) \surround{m\leq 3}{hops} \: \device$. }
\label{fig:spprop}
}
\end{figure}

We can also derive  other three spatial operators:  {\it somewhere},  {\it everywhere} and  {\it surround}. 

\noindent{\bf Somewhere.}  $ \somewhere{ d }{f} \varphi := true  \reach{ d}{f} \varphi $
is satisfied by $(\lambda , x(t,\ell))$
iff there exists a location that satisfies $\varphi$ reachable from $\ell$ via a route $\tau$ with a distance  satisfying the predicate $\delta(d)$. This length is computed via the function $\gamma(f)$. In Fig.~\ref{fig:spprop}, all the locations  satisfy the property $\somewhere{m \leq 4 }{hops} \coord$ because, for all $\ell_i$, there is always a 
path $\route = \ell_i \dots \ell_{10}$ with a length $d_\route^{hops}[k]\leq 4$, where $\tau[0]=\ell_{i}$, $\tau[k]=\ell_{10}$, and $\ell_{10}$ satisfies the green property, i.e. it is a coordinator node.

\noindent{\bf Everywhere.}  $ \everywhere{d}{f} \varphi := \neg \somewhere{d}{f} \neg \varphi $ 
is satisfied by $(\lambda , x(t,\ell))$ iff  all the locations reachable from $\ell$ via a path, with length satisfying the predicate  $\delta(d)$, satisfy $\varphi$. In Fig.~\ref{fig:spprop}, there are no locations that satisfy the property $\everywhere{m \leq 2 }{hops} \router$ because for all the locations $\ell_i$ there is a path $\tau=\ell_i\ell_j$ s.t. $\ell_j$ is not a router.

\noindent{\bf {Surround}.}  $\varphi_{1} \surround{ d }{f} \varphi_{2} := \varphi_{1} \wedge \neg (\varphi_{1}\reach{ d}{f} \neg (\varphi_1 \vee \varphi_{2}) \wedge \neg (\escape{\neg d}{f}  \varphi_{1}) $
expresses the topological notion of being surrounded by a $\varphi_2$-region, while being in a $\varphi_{1}$-region,  with an additional metric constraint. The operator has been introduced in~\cite{CianciaLLM16} as a basic operator, while here it is a derived one. The idea is that one cannot escape from a $\varphi_{1}$-region without passing from a location that satisfies $\varphi_2$ and, in any case, one has to reach a $\varphi_2$-location via a path with a length satisfying the predicate $d$.  In Fig.~\ref{fig:spprop}, the location $\ell_{10}$ satisfies the property $ (\coord  \: \vee  \: \router ) \surround{\leq 3}{hops} \: \device$. In fact, it is coordinator,  it cannot reach a location that does not satisfy the  the $\coord  \: \vee  \: \router$ or the $\device$ property via a path with length lesser or equal to 3 and it cannot escape through a path satisfying the $\coord  \: \vee  \: \router$ property at a distance more than 3. 

The operators can be arbitrarily composed to specify complex properties as we will see in Section~\ref{sec:results}. Furthermore, they can be evaluated both on indirect and on direct graphs.


\section{Monitoring STREL}
\label{sec:alg}
In this section, we present a 
monitoring algorithm that can be used to check if a given 
signal satisfies or not a STREL property. 
The proposed algorithm follows an \emph{offline} approach. Indeed, it takes as input the complete spatio-temporal signal together with the property we want to monitor. 
 At the end of this section, we will also briefly discuss a possible alternative approach that can lead to a distributed and \emph{online} monitoring procedure.
In this case, the spatio-temporal signal is not known at the beginning, it is discovered while data are collected from the system during its execution.
%
%


\subsection{Offline monitor}
Offline monitoring is performed via the function $\mathsf{monitor}$ 
that takes as inputs a location service $\lserv$, a trace $\vec{x}$ 
and a formula $\phi$ and returns the \emph{piecewise constant 
spatio-temporal signal} $\pcsts$ representing the monitoring 
of $\phi$.
The function also relies on parametrised with respect to functions $\iota$, $\delta$ and $\gamma$, used to interpret symbols
in formulas, and  operators $\oplus_{D_2}$, $\otimes_{D_2}$ and $\odot_{D_{2}}$ of \emph{signal domain}, used to represent 
satisfaction values.

The function $\mathsf{monitor}$ is defined by induction on the syntax of the formula\footnote{This definition is straightforward and, for the sake of readability, we only report it in Appendix, available in the extend version of this article at  \url{https://github.com/Quanticol/strel}}. 
The spatio-temporal signal resulting from the monitoring of atomic proposition $\mu$ is just obtained by applying function $\iota(\mu)$ to the trace $\mathbf{x}$. The spatio-temporal signals associated with $\neg\varphi$ and $\varphi_1\wedge \varphi_2$ are obtained by applying operators $\odot_{D_2}$ and $\otimes_{D_2}$ to the signals resulting from the monitoring of $\varphi$ and from the monitoring of $\varphi_1$ and $\varphi_2$. 

Monitoring of temporal properties, namely $\varphi_1 \until{\leq t}\varphi_2$ and $\varphi_1 \since{\leq t} \varphi_2$, can be done by using the same approach used in~\cite{Donze2013} and~\cite{MalerN13}. However, while their monitoring relies on classical boolean and arithmetic operators, here the procedure is parametrised with respect to operators $\oplus_{D_2}$ and $\otimes_{D_2}$ of the considered semiring.  

To monitor $\varphi_1\reach{d}{f}\varphi_2$ first the signals $\mathbf{s}_1$ and $\mathbf{s}_2$ resulting from the monitoring of $\varphi_1$ and $\varphi_2$ are computed. After that, the final result is computed by aggregating the spatial signals $\mathbf{s}_1@t$ and $\mathbf{s}_2@t$ at each time $t\in \mathcal{T}(\mathbf{s}_1)\cup \mathcal{T}(\mathbf{s}_2)$ with  function $\mathsf{reach}$, defined in Algoritm~\ref{algo:reachmonitoring}.
This function also takes as parameters the spatial model $(L,\wfun)$ at time $t$ (obtained from the \emph{location service}), the function $f:B\times A\rightarrow B$ used to compute the distances over paths, and the predicate $d$ describing the reachability bound.
In function $\mathsf{reach}$, the data structure $r$ is iteratively computed. This data structure associates each location $\ell$ with a set of triples $(\ell', v, w)$. Intuitively, $(\ell', v, w)$ is in $r[\ell]$ after $i$ iterations if and only if: $\ell$ can reach $\ell'$ with at most $i$-steps with a distance at least $w$ ($w$ satisfying $d$) and a monitored value $v$. 
At the beginning $r[\ell]$ is initialised to $\{ (\ell, \ssign_2(\ell), 0) \}$. Moreover, at each iteration, the values in $r[\ell]$ are updated by considering the elements in $r[\ell']$, for any $\ell'$ next to $\ell$.
The loop continues until a fix point is reached. 
Note that, termination of the algorithm is guaranteed by the fact that $D_2$ is an \emph{idempotent semiring} and from the fact that, for any $(\ell_1,v_1,w_1), (\ell_2,v_2,w_2) \in r[\ell]$, if $\ell_1=\ell_2$ and $v_1=v_2$ then $w_1=w_2$.
The result spatial signal associates each location $\ell$ with the value $\bigoplus_{D_2}(\{ v | (\ell',v,w)\in r[\ell] \})$.

Monitoring algorithm for $\escape{d}{f} \varphi$ is reported in Algorithm~\ref{algo:escapemonitoring}, where function $\mathsf{escape}$ is defined. Given a space model at time $t$, a distance function $f$, a distance predicate $d$ and a spatial signal, it computes the spatial signal representing the monitoring value of $\escape{d}{f} \varphi$ at time $t$. 
%
Function $\mathsf{escape}$ iteratively computes the data structure obtained by $e$ that associates each location $\ell$ with a set of triples of the form $(\ell',v,w)$ representing the fact that $\ell$ can \emph{escape} in $\ell'$ with a distance $w$ and a total value $v$. At each iteration, these values are updated by considering the values in the neighbours in each location. 
Similarly to function $\mathsf{reach}$, this computation continues until a fixpoint is reached. After that, the monitored value associated with each location $\ell$ is computed as $\bigoplus_{D_2}(\{ v | (\ell',v,w)\in e[\ell] \wedge d(w) \})$.


\begin{algorithm}[tbp] 
\caption{Function $\mathsf{reach}$}
\label{algo:reachmonitoring}
\vspace{1mm}
\begin{algorithmic}[1]
\State inputs: $(L,\wfun)$, $f:B\times A\rightarrow B$, $d: B\rightarrow \{ true,false\}$, $\ssign_1$, $\ssign_2$ 
\State $\forall \ell\in L. r[\ell] = \{ (\ell, \ssign_2(\ell), 0) \}$
\State $stable = false$
\While{ $\neg stable$ }
\State $stable = true$
\State $r'=r$ 
\ForAll{ $\ell_1\in L$ }
\ForAll{ $\ell_2: \nextto{\ell_1}{w}{\ell_2}$ }
\State{ \small$N=\{ (\ell, v\otimes_{D_2} \ssign_1(\ell_1), f(w',w)) | (\ell,v, w')\in r[\ell_2] \wedge d(f(w',w))\}$}
\ForAll{ $(\ell,v,w) \in N$}
\If{$\exists (\ell,v,w')\in r'[\ell]$}
\State{\footnotesize $r'[\ell]=r'[\ell]-\{ (\ell,v,w') \} \cup \{ (\ell,v,w\oplus_{B} w') \}$}
\Else
\State{ \footnotesize  $r'[\ell]=r'[\ell] \cup \{ (\ell,v,w) \}$}
\EndIf
\EndFor
\EndFor
\If{ $r'[\ell] \not= r[\ell]$ }
\State $stable = false$
\EndIf
\EndFor
\State $r=r'$
\EndWhile
\State $\ssign=[]$
\ForAll{ $\ell\in L$ }
\State $\ssign(\ell)=\bigoplus_{D_2}(\{ v | (\ell',v,w)\in r[\ell] \})$
\EndFor{}
\State \Return{ $\ssign$}
\end{algorithmic}
\end{algorithm}

\begin{algorithm}[tbp] 
\caption{Function $\mathsf{escape}$}
\label{algo:escapemonitoring}
\vspace{1mm}
\begin{algorithmic}[1]
\State inputs: $(L,\wfun)$,$f:B\times A\rightarrow B$, $d: B\rightarrow \{ true,false\}$,$\ssign_1$ 
\State $\forall \ell\in L. e[\ell] = \{ (\ell, \ssign_1(\ell),0) \}$
\State $stable = false$
\While{ $\neg stable$ }
\State $stable = true$
\State $e'=e$ 
\ForAll{ $\ell_1\in L$ }
\ForAll{ $\ell_2: \nextto{\ell_1}{w}{\ell_2}$ }
\State{ \small $N=\{ (\ell, v\otimes_{D_2} \ssign_1(\ell_1), f(w',w)) | (\ell,v,w')\in r[\ell_2]\}$}
\ForAll{ $(\ell,v,w) \in N$}
\If{$\exists (\ell,v',w')\in e'[\ell_1]$}
\State{\small $e'[\ell_1]=e'[\ell_1]-\{ (\ell,v',w') \} \cup \{ (\ell,v\oplus_{D_2} v',w\oplus_{B} w') \}$}
\Else
\State $e'[\ell]=e '[\ell] \cup \{ (\ell,v,w) \}$
\EndIf
\EndFor
\EndFor
\If{ $e'[\ell] \not= e[\ell]$ }
\State $stable = false$
\EndIf
\EndFor
\State $e=e'$
\EndWhile
\State $\ssign=[]$
\ForAll{ $\ell\in L$ }
\State $\ssign(\ell)=\bigoplus_{D_2}(\{ v | (\ell',v,w)\in e[\ell] \wedge d(w) \})$
\EndFor{}
\State \Return $\ssign$
\end{algorithmic}
\end{algorithm}

\noindent {\bf Remark.} The offline monitoring iteratively computes the monitor value at a location by considering the values of monitoring in the previous iteration. This approach easily enables the definition of a parallel monitoring algorithm. Indeed, both the functions defined in Algorithm~\ref{algo:reachmonitoring} and Algorithm~\ref{algo:escapemonitoring} can be parallel executed for each location. The different monitoring instances must communicate to exchange the values computed at iteration $i$. 
Another possible improvement of this algorithm is based on an \emph{online} computation of the monitoring. Following an approach similar to the one considered in~\cite{DDGJJS15}, each location can identify its monitoring value by using only partial informations. Early termination of the monitor procedure is then possible when the satisfaction or violation of a property is found.


\section{Examples}
\label{sec:results}

In this section we present some example of the expressibility and potentiality of STREL. 

\subsection{ZigBee protocol monitoring}
\newcommand{\nM}{50}
\newcommand{\nVar}{3}

Given a MANET with a ZigBee protocol (Example~\ref{ex:zigbee}), 
we consider as spatial models both its proximity and connectivity graphs, computed with respect to the Cartesian coordinates. The Nodes have three kinds of roles: {\it coordinator}, {\it router} and {\it EndDevice}, as described in Example \ref{ex:zigbee}. Moreover, each device is also equipped with a sensor to monitor its battery level ($X_B$), the humidity ($X_H$) and the pollution  ($X_H$) in its position.  
The semiring is the union between the  \emph{max/min} semiring $\mathbb{R}^{\infty}$ (for the proximity graph) and the  \emph{integer} semiring $\mathbb{N}^{\infty}$ (for the connectivity graph). We will use also two types of distances: ${\it hops}$ and the $\Delta$ distances described in Example~\ref{ex:distancefunction}.
As in the Example~\ref{ex:zigbee}, atomic propositions  $\{ \coord, \router, \device\}$ describe the type of nodes. We also consider inequalities on the values that are read from sensors, plus special propositions $@_\ell$ which encode the address of a specific location, i.e. they are true only in the location $\ell$. 

In the following, we describe several properties of these ZigBee MANET networks that are easily captured by STREL logic, to exemplify its expressive power.

A class of properties naturally encoded in STREL related to the connectivity of the network. First, we can be interested to know if a node is properly connected, meaning that it can reach the coordinator through a path of routers:
\begin{equation}
\phi_{connect} = \device \reach{m \leq 1}{hops} (\router \reach{ m < \infty}{hops} \coord )
\end{equation}
The meaning of this property is that an end node reaches in a step a node which is a router and that is connected to the coordinator via a path of routers.

We may also want to know if there is a path to the router which is reliable in terms of battery levels, for instance such that all routers have a battery level above 30\%:  
\begin{eqnarray}
&\phi_{reliable\_router} = ((X_{B} > 30\%) \wedge \router) \reach{ m < \infty}{hops} \coord &\nonumber \\
&\phi_{reliable\_connect} =   \device \reach{m \leq 1}{hops} (\phi_{reliable\_router}  )&
\end{eqnarray}
The properties focus on spatial connectivity at a fixed time. We can add also temporal requirements, for instance asking that a broken connection is restored within $h$ time units:
\begin{equation}
\phi_{connect\_restore} = \glob{} (\neg \phi_{connect} \rightarrow  \ev{[0,h]}\phi_{connect} )
\end{equation}
Another class of properties of interest is  the acyclicity of transmissions. To this end, we need to force the connectivity graph to be direct, with edges pointing in the direction of the coordinator (i.e. transmission reduces the distance from the coordinator).  With STREL, we can easily detect the  absence of a cycle locally, i.e. for a fixed location $\ell$. This is captured by 
$\phi^{\ell}_{acyclic} = \neg \phi^{\ell}_{cycle}$, where 
\begin{equation}
\phi^{\ell}_{acyclic} =  @_\ell \reach{m \leq 1}{hops}  (\neg  @_\ell  \wedge \somewhere{}{hops}@_\ell)
\end{equation}
In order to characterize the whole network as acyclic, we need to take the conjunction of the previous formulae for all locations (or at least for routers, enforcing end devices to be connected only with routers). This is necessary as STREL is interpreted locally, on each location, and this forbids us to express properties of the whole network with location unaware formulae. This is a price for an efficient monitoring, as global properties of networks require more expressive and computationally expensive logics. 
However, we can use the parametrization of STREL and the property of a Voronoi diagram to specify the global connection or the acyclicity of the graph. Indeed, the proximity graph connects always all the locations of the system, then the property $\everywhere{}{\Delta} \phi$, verified on the proximity graph, holds iff $\phi$ holds in all the location of the system.
 
Up to now we have presented qualitative properties, depending on the type of node. If we express properties of sensor measurements, we can also consider a quantitative semantics, returning a measure of robustness of (dis)satisfaction. As an example, we  can monitor \eqref{eq:f1} if in each location an high value of pollution eventually implies,  within $T$ time units, an high value of humidity, or \eqref{eq:f2} in which locations it is possible to find a `safe' route, where both the humidity and the pollution are below a certain threshold. We can also check \eqref{eq:f3} if a location, which is not safe, is at distance at most $5$ from a location which is safe. Finally \eqref{eq:f4}, we can check if a target device (identified by $X_S=1$) is reachable from all the locations in less than 10 hops.
\begin{eqnarray}
&\phi_{PH} = (X_P > 150) \Rightarrow \ev{[0,T]} (X_H > 100) \label{eq:f1} &\\
&\phi_{Safe} =\glob{[0,T]} \escape{m \geq k}{\Delta} \: {(X_H < 90) \wedge (X_P < 150) } \label{eq:f2} &\\
&\phi_{some} = \somewhere{m \leq 5}{\Delta}  \phi_{Safe} \label{eq:f3} &\\
&\phi_{target} = \everywhere{}{hops} \somewhere{m < 10}{hops}\: { (X_S = 1) }  \label{eq:f4} &
\end{eqnarray}

\subsection{Invariance properties of the Euclidean spatial model}

The properties we consider with respect to the Euclidean spatial model are typically local and depend on the relative distance and position among nodes in the plane. As such, they should be invariant with respect to change of coordinates, i.e. with respect to isometric transformations of the plane. This class of transformations includes translations, rotations, and reflections, and can be described by matrix multiplications of the form   
\[
\begin{bmatrix}
    x'_{\ell}      \\
    y'_{\ell}    \\ 
    1              \\ 
\end{bmatrix}
= 
\begin{bmatrix}
       \beta \cos (\alpha)   &   -   \beta \sin (\alpha)  & \beta t_x     \\
       \gamma \sin (\alpha)  &       \gamma \cos (\alpha)  & \gamma t_y \\
    0 &  0 & 1     
\end{bmatrix} 
\begin{bmatrix}
    x_{\ell}      \\
    y_{\ell}    \\ 
    1              \\ 
\end{bmatrix}
\]

Invariance of satisfaction of spatial properties holds in STREL logic, for the Euclidean space model of Definition \ref{def:Euclidean}. Consider more specifically an Euclidean space model $\mathcal{E}(L,\mu, R) = \langle L, \wfun^{\mu, R} \rangle$ and $\mathcal{E}(L,\mu', R)= \langle L, \wfun^{\mu', R} \rangle$, obtained by applying an isometric transformation $A$:  $\mu'(\ell) = A(\mu(\ell))$. For invariance to hold, we need to further require that distance predicates, used in spatial operators, are invariant for isometric transformations. More specifically, for any isometry $A$, we require a distance predicate $d$ on the semiring $\mathbb{R}^{\infty}\times\mathbb{R}^{\infty}$ to satisfy $d((x,y)) = d(A((x,y)))$. This is the case for the norm-based predicates used in the examples,  of the form $d((x,y)) = \|(x,y\|_2 \leq r$.  

Notice that, the path structure is preserved (the edges given by $R$ is the same), and the truth of isometry-invariant distance predicates along paths in $\mathcal{E}(L,\mu, R)$ and $\mathcal{E}(L,\mu', R)$ is also the same. This straightforwardly implies that the truth value of spatial operators will be unchanged by isometry. 

\begin{proposition}[Equisatisfiability under isometry] Let  $\mathcal{E}(L,\mu, R) = \langle L, \wfun^{\mu, R} \rangle$
be  an euclidean spatial model and  $\mathcal{E}(L,\mu', R)= \langle L, \wfun^{\mu', R} \rangle$ an isometric transformation of the former. Consider a spatial formula $\varphi_{1} \: \reach{ d}{f} \: \varphi_{2}$ or $\escape{d}{f}  \: \varphi_{1}$, where $d$ is an isometry preserving predicate.
Assume $\fmon ( \lserv, \vec{x}, \varphi_{j}, t, \ell) = \fmon'( \lserv, \vec{x}, \varphi_{j}, t, \ell)$, $j=1,2$, where $\fmon$ and $\fmon'$ are the monitoring functions for the two spatial models. Then it holds that 
$\fmon( \lserv, \vec{x}, \varphi_{1} \: \reach{ d}{f} \: \varphi_{2}, t, \ell) = \fmon'( \lserv, \vec{x}, \varphi_{1} \: \reach{ d}{f} \: \varphi_{2}, t, \ell)$ and $\fmon(\lserv, \vec{x}, \escape{d}{f}  \: \varphi_{1}, t, \ell) = \fmon'( \lserv, \vec{x}, \escape{d}{f}  \: \varphi_{1}, t, \ell)$, for all $\ell$ and $t$. 
\end{proposition}

\section{Conclusion and Future Work}
\label{sec:conclusion}

The rise of mobile and spatially distributed CPS demands 
for novel efficient and effective spatio-temporal formal 
frameworks to specify concisely spatio-temporal requirements 
and to enable the qualitative and quantitative spatio-temporal 
monitoring of such properties over spatially distributed CPS.
STREL provides an intuitive formal framework that 
enable to express formally spatio-temporal requirements 
and to monitor them automatically over the execution of \emph{mobile} 
and \emph{spatially distributed} CPS.  We have demonstrated 
the feasibility of our approach showing an application of
STREL to monitor a simulated mobile ad hoc sensor network.
While in this paper we define the logic and provide an offline monitoring algorithm, 
future research includes the design of distributed monitoring
algorithms, a thorough investigation of the expressiveness, learning STREL requirements directly from data 
and synthesizing control policies to ensure a given requirement.
A set of API that implements the algorithms considered in this paper is currently under development\footnote{STREL API are public available at \url{https://github.com/Quanticol/strel}}.

\section*{Acknowledgment}

L.B.\, L.N.\ and M.L.\ acknowledge partial support from the 
EU-FET project QUANTICOL (nr. 600708).
E.B.\ and L.N.\ acknowledge the partial support of the Austrian National 
Research Network  S 11405-N23 (RiSE/SHiNE) of the Austrian Science 
Fund (FWF), the ICT COST Action IC1402 Runtime Verification beyond 
Monitoring (ARVI).


\bibliographystyle{ACM-Reference-Format}
\bibliography{biblio} 


\begin{thebibliography}{00}


\ifx \showCODEN    \undefined \def \showCODEN     #1{\unskip}     \fi
\ifx \showDOI      \undefined \def \showDOI       #1{#1}\fi
\ifx \showISBNx    \undefined \def \showISBNx     #1{\unskip}     \fi
\ifx \showISBNxiii \undefined \def \showISBNxiii  #1{\unskip}     \fi
\ifx \showISSN     \undefined \def \showISSN      #1{\unskip}     \fi
\ifx \showLCCN     \undefined \def \showLCCN      #1{\unskip}     \fi
\ifx \shownote     \undefined \def \shownote      #1{#1}          \fi
\ifx \showarticletitle \undefined \def \showarticletitle #1{#1}   \fi
\ifx \showURL      \undefined \def \showURL       {\relax}        \fi
\providecommand\bibfield[2]{#2}
\providecommand\bibinfo[2]{#2}
\providecommand\natexlab[1]{#1}
\providecommand\showeprint[2][]{arXiv:#2}

\bibitem[\protect\citeauthoryear{Aurenhammer}{Aurenhammer}{1991}]%
        {Aurenhammer1991}
\bibfield{author}{\bibinfo{person}{F. Aurenhammer}.}
  \bibinfo{year}{1991}\natexlab{}.
\newblock \showarticletitle{Voronoi Diagrams; a Survey of a Fundamental
  Geometric Data Structure}.
\newblock \bibinfo{journal}{{\em ACM Comput. Surv.\/}} \bibinfo{volume}{23},
  \bibinfo{number}{3} (\bibinfo{year}{1991}), \bibinfo{pages}{345--405}.
\newblock
\showISSN{0360-0300}
\showDOI{%
\url{https://doi.org/10.1145/116873.116880}}


\bibitem[\protect\citeauthoryear{Aydin-Gol, Bartocci, and Belta}{Aydin-Gol
  et~al\mbox{.}}{2014}]%
        {bartocci2014}
\bibfield{author}{\bibinfo{person}{A. Aydin-Gol}, \bibinfo{person}{E.
  Bartocci}, {and} \bibinfo{person}{C. Belta}.}
  \bibinfo{year}{2014}\natexlab{}.
\newblock \showarticletitle{A Formal Methods Approach to Pattern Synthesis in
  Reaction Diffusion Systems}. In \bibinfo{booktitle}{{\em Proc. of CDC: the
  53rd {IEEE} Conference on Decision and Control}}.
  \bibinfo{publisher}{{IEEE}}, \bibinfo{pages}{108--113}.
\newblock
\showDOI{%
\url{https://doi.org/10.1109/CDC.2014.7039367}}


\bibitem[\protect\citeauthoryear{Bartocci, Aydin-Gol, Haghighi, and
  Belta}{Bartocci et~al\mbox{.}}{2016}]%
        {Bartocci2016}
\bibfield{author}{\bibinfo{person}{E. Bartocci}, \bibinfo{person}{E.
  Aydin-Gol}, \bibinfo{person}{I. Haghighi}, {and} \bibinfo{person}{C. Belta}.}
  \bibinfo{year}{2016}\natexlab{}.
\newblock \showarticletitle{A Formal Methods Approach to Pattern Recognition
  and Synthesis in Reaction Diffusion Networks}.
\newblock \bibinfo{journal}{{\em IEEE Transactions on Control of Network
  Systems\/}} \bibinfo{volume}{PP}, \bibinfo{number}{99}
  (\bibinfo{year}{2016}), \bibinfo{pages}{1--1}.
\newblock
\showISSN{2325-5870}
\showDOI{%
\url{https://doi.org/10.1109/TCNS.2016.2609138}}


\bibitem[\protect\citeauthoryear{Bennett, Cohn, Wolter, and
  Zakharyaschev}{Bennett et~al\mbox{.}}{2002}]%
        {BC02}
\bibfield{author}{\bibinfo{person}{B. Bennett}, \bibinfo{person}{A.~G. Cohn},
  \bibinfo{person}{F. Wolter}, {and} \bibinfo{person}{M. Zakharyaschev}.}
  \bibinfo{year}{2002}\natexlab{}.
\newblock \showarticletitle{Multi-Dimensional Modal Logic As a Framework for
  Spatio-Temporal Reasoning}.
\newblock \bibinfo{journal}{{\em Applied Intelligence\/}} \bibinfo{volume}{17},
  \bibinfo{number}{3} (\bibinfo{date}{Sept.} \bibinfo{year}{2002}),
  \bibinfo{pages}{239--251}.
\newblock
\showISSN{0924-669X}
\showDOI{%
\url{https://doi.org/10.1023/A:1020083231504}}


\bibitem[\protect\citeauthoryear{Bistarelli, Montanari, and Rossi}{Bistarelli
  et~al\mbox{.}}{1997}]%
        {BMR97}
\bibfield{author}{\bibinfo{person}{S. Bistarelli}, \bibinfo{person}{U.
  Montanari}, {and} \bibinfo{person}{F. Rossi}.}
  \bibinfo{year}{1997}\natexlab{}.
\newblock \showarticletitle{Semiring-based constraint satisfaction and
  optimization}.
\newblock \bibinfo{journal}{{\em J. {ACM}\/}} \bibinfo{volume}{44},
  \bibinfo{number}{2} (\bibinfo{year}{1997}), \bibinfo{pages}{201--236}.
\newblock
\showDOI{%
\url{https://doi.org/10.1145/256303.256306}}


\bibitem[\protect\citeauthoryear{Bresolin, Sala, Della~Monica, Montanari, and
  Sciavicco}{Bresolin et~al\mbox{.}}{2010}]%
        {BS10}
\bibfield{author}{\bibinfo{person}{D. Bresolin}, \bibinfo{person}{P. Sala},
  \bibinfo{person}{D. Della~Monica}, \bibinfo{person}{A. Montanari}, {and}
  \bibinfo{person}{G. Sciavicco}.} \bibinfo{year}{2010}\natexlab{}.
\newblock \showarticletitle{A Decidable Spatial Generalization of Metric
  Interval Temporal Logic}. In \bibinfo{booktitle}{{\em Proc. of TIME 2010: the
  17th International Symposium on Temporal Representation and Reasoning}}.
  \bibinfo{pages}{95--102}.
\newblock
\showISSN{1530-1311}
\showDOI{%
\url{https://doi.org/10.1109/TIME.2010.22}}


\bibitem[\protect\citeauthoryear{Caires and Cardelli}{Caires and
  Cardelli}{2003}]%
        {CC04}
\bibfield{author}{\bibinfo{person}{L. Caires} {and} \bibinfo{person}{L.
  Cardelli}.} \bibinfo{year}{2003}\natexlab{}.
\newblock \showarticletitle{A spatial logic for concurrency (part I)}.
\newblock \bibinfo{journal}{{\em Information and Computation\/}}
  \bibinfo{volume}{186}, \bibinfo{number}{2} (\bibinfo{year}{2003}),
  \bibinfo{pages}{194--235}.
\newblock
\showISSN{0890-5401}
\showDOI{%
\url{https://doi.org/10.1016/S0890-5401(03)00137-8}}


\bibitem[\protect\citeauthoryear{Ciancia, Latella, Loreti, and Massink}{Ciancia
  et~al\mbox{.}}{2016}]%
        {CianciaLLM16}
\bibfield{author}{\bibinfo{person}{V. Ciancia}, \bibinfo{person}{D. Latella},
  \bibinfo{person}{M. Loreti}, {and} \bibinfo{person}{M. Massink}.}
  \bibinfo{year}{2016}\natexlab{}.
\newblock \showarticletitle{Spatial Logic and Spatial Model Checking for
  Closure Spaces}. In \bibinfo{booktitle}{{\em {SFM} 2016: 16th Intern. School
  on Formal Methods for the Design of Computer, Communication, and Software
  Systems}} {\em (\bibinfo{series}{LNCS})}, Vol.~\bibinfo{volume}{9700}.
  \bibinfo{publisher}{Springer}, \bibinfo{pages}{156--201}.
\newblock
\showDOI{%
\url{https://doi.org/10.1007/978-3-319-34096-8_6}}


\bibitem[\protect\citeauthoryear{Delaunay}{Delaunay}{1934}]%
        {Delaunay1934}
\bibfield{author}{\bibinfo{person}{B. Delaunay}.}
  \bibinfo{year}{1934}\natexlab{}.
\newblock \showarticletitle{Sur la sph\'ere vide}.
\newblock \bibinfo{journal}{{\em Bulletin de l'Acad\'emie des Sciences de
  l'URSS, Classe des sciences math\'ematiques et naturelles\/}}
  \bibinfo{volume}{6} (\bibinfo{year}{1934}), \bibinfo{pages}{793--800}.
\newblock


\bibitem[\protect\citeauthoryear{Deshmukh, Donz\'{e}, Ghosh, Jin, Juniwal, and
  Seshia}{Deshmukh et~al\mbox{.}}{2015}]%
        {DDGJJS15}
\bibfield{author}{\bibinfo{person}{J.~V. Deshmukh}, \bibinfo{person}{A.
  Donz\'{e}}, \bibinfo{person}{S. Ghosh}, \bibinfo{person}{X. Jin},
  \bibinfo{person}{G. Juniwal}, {and} \bibinfo{person}{S.~A. Seshia}.}
  \bibinfo{year}{2015}\natexlab{}.
\newblock \showarticletitle{Robust Online Monitoring of Signal Temporal Logic}.
  In \bibinfo{booktitle}{{\em Proc. of {RV} 2015: the 6th International
  Conference on Runtime Verification}} {\em (\bibinfo{series}{LNCS})},
  Vol.~\bibinfo{volume}{9333}. \bibinfo{publisher}{Springer},
  \bibinfo{pages}{55--70}.
\newblock
\showDOI{%
\url{https://doi.org/10.1007/978-3-319-23820-3_4}}


\bibitem[\protect\citeauthoryear{Donz{\'e}, Ferrer, and Maler}{Donz{\'e}
  et~al\mbox{.}}{2013}]%
        {Donze2013}
\bibfield{author}{\bibinfo{person}{A. Donz{\'e}}, \bibinfo{person}{T. Ferrer},
  {and} \bibinfo{person}{O. Maler}.} \bibinfo{year}{2013}\natexlab{}.
\newblock \showarticletitle{Efficient Robust Monitoring for STL}. In
  \bibinfo{booktitle}{{\em Proc. of CAV 2013: the 25th International Conference
  on Computer Aided Verification}} {\em (\bibinfo{series}{LNCS})}.
  \bibinfo{pages}{264--279}.
\newblock
\showDOI{%
\url{https://doi.org/10.1007/978-3-642-39799-8_19}}


\bibitem[\protect\citeauthoryear{Grosu, Smolka, Corradini, Wasilewska,
  Entcheva, and Bartocci}{Grosu et~al\mbox{.}}{2009}]%
        {GrosuSCWEB09}
\bibfield{author}{\bibinfo{person}{R. Grosu}, \bibinfo{person}{S.~A. Smolka},
  \bibinfo{person}{F. Corradini}, \bibinfo{person}{A. Wasilewska},
  \bibinfo{person}{E. Entcheva}, {and} \bibinfo{person}{E. Bartocci}.}
  \bibinfo{year}{2009}\natexlab{}.
\newblock \showarticletitle{Learning and detecting emergent behavior in
  networks of cardiac myocytes}.
\newblock \bibinfo{journal}{{\em Commun. {ACM}\/}} \bibinfo{volume}{52},
  \bibinfo{number}{3} (\bibinfo{year}{2009}), \bibinfo{pages}{97--105}.
\newblock
\showDOI{%
\url{https://doi.org/10.1145/1467247.1467271}}


\bibitem[\protect\citeauthoryear{Haghighi, Jones, Kong, Bartocci, R., and
  Belta}{Haghighi et~al\mbox{.}}{2015}]%
        {bartocci2015}
\bibfield{author}{\bibinfo{person}{I. Haghighi}, \bibinfo{person}{A. Jones},
  \bibinfo{person}{J.~Z. Kong}, \bibinfo{person}{E. Bartocci},
  \bibinfo{person}{Grosu R.}, {and} \bibinfo{person}{C. Belta}.}
  \bibinfo{year}{2015}\natexlab{}.
\newblock \showarticletitle{{SpaTeL}: {A} {Novel} {Spatial}-{Temporal} {Logic}
  and {Its} {Applications} to {Networked} {Systems}}. In
  \bibinfo{booktitle}{{\em Proc. of HSCC 2015: the 18th International
  Conference on Hybrid Systems: Computation and Control}}.
  \bibinfo{publisher}{{ACM}}, \bibinfo{pages}{189--198}.
\newblock
\showDOI{%
\url{https://doi.org/10.1145/2728606.2728633}}


\bibitem[\protect\citeauthoryear{Lluch{-}Lafuente and
  Montanari}{Lluch{-}Lafuente and Montanari}{2005}]%
        {LM05}
\bibfield{author}{\bibinfo{person}{A. Lluch{-}Lafuente} {and}
  \bibinfo{person}{U. Montanari}.} \bibinfo{year}{2005}\natexlab{}.
\newblock \showarticletitle{Quantitative mu-calculus and {CTL} defined over
  constraint semirings}.
\newblock \bibinfo{journal}{{\em Theor. Comput. Sci.\/}} \bibinfo{volume}{346},
  \bibinfo{number}{1} (\bibinfo{year}{2005}), \bibinfo{pages}{135--160}.
\newblock
\showDOI{%
\url{https://doi.org/10.1016/j.tcs.2005.08.006}}


\bibitem[\protect\citeauthoryear{Maler and Nickovic}{Maler and
  Nickovic}{2013}]%
        {MalerN13}
\bibfield{author}{\bibinfo{person}{O. Maler} {and} \bibinfo{person}{D.
  Nickovic}.} \bibinfo{year}{2013}\natexlab{}.
\newblock \showarticletitle{Monitoring properties of analog and mixed-signal
  circuits}.
\newblock \bibinfo{journal}{{\em {STTT}\/}} \bibinfo{volume}{15},
  \bibinfo{number}{3} (\bibinfo{year}{2013}), \bibinfo{pages}{247--268}.
\newblock
\showDOI{%
\url{https://doi.org/10.1007/s10009-012-0247-9}}


\bibitem[\protect\citeauthoryear{Marx and Reynolds}{Marx and Reynolds}{1999}]%
        {MR99}
\bibfield{author}{\bibinfo{person}{M. Marx} {and} \bibinfo{person}{M.
  Reynolds}.} \bibinfo{year}{1999}\natexlab{}.
\newblock \showarticletitle{Undecidability of compass logic}.
\newblock \bibinfo{journal}{{\em J Logic Computation\/}} \bibinfo{volume}{9},
  \bibinfo{number}{6} (\bibinfo{year}{1999}), \bibinfo{pages}{897--914}.
\newblock
\showDOI{%
\url{https://doi.org/10.1093/logcom/9.6.897}}


\bibitem[\protect\citeauthoryear{Mottola, Moretta, Whitehouse, and
  Ghezzi}{Mottola et~al\mbox{.}}{2014}]%
        {MottolaMWG14}
\bibfield{author}{\bibinfo{person}{L. Mottola}, \bibinfo{person}{M. Moretta},
  \bibinfo{person}{K. Whitehouse}, {and} \bibinfo{person}{C. Ghezzi}.}
  \bibinfo{year}{2014}\natexlab{}.
\newblock \showarticletitle{Team-level programming of drone sensor networks}.
  In \bibinfo{booktitle}{{\em Proc. of the 12th {ACM} Conference on Embedded
  Network Sensor Systems, SenSys '14, Memphis, Tennessee, USA, November 3-6,
  2014}}. \bibinfo{publisher}{{ACM}}, \bibinfo{pages}{177--190}.
\newblock
\showDOI{%
\url{https://doi.org/10.1145/2668332.2668353}}


\bibitem[\protect\citeauthoryear{Nenzi, Bortolussi, Ciancia, Loreti, and
  Massink}{Nenzi et~al\mbox{.}}{2015}]%
        {NenziBCLM15}
\bibfield{author}{\bibinfo{person}{L. Nenzi}, \bibinfo{person}{L. Bortolussi},
  \bibinfo{person}{V. Ciancia}, \bibinfo{person}{M. Loreti}, {and}
  \bibinfo{person}{M. Massink}.} \bibinfo{year}{2015}\natexlab{}.
\newblock \showarticletitle{Qualitative and Quantitative Monitoring of
  Spatio-Temporal Properties}. In \bibinfo{booktitle}{{\em Proc. of RV 2015:
  the 6th International Conference on Runtime Verification}} {\em
  (\bibinfo{series}{LNCS})}, Vol.~\bibinfo{volume}{9333}.
  \bibinfo{publisher}{Springer}, \bibinfo{pages}{21--37}.
\newblock
\showDOI{%
\url{https://doi.org/10.1007/978-3-319-23820-3_2}}


\bibitem[\protect\citeauthoryear{Talcott}{Talcott}{2008}]%
        {Talcott08}
\bibfield{author}{\bibinfo{person}{C.~L. Talcott}.}
  \bibinfo{year}{2008}\natexlab{}.
\newblock \showarticletitle{Cyber-Physical Systems and Events}.
\newblock In \bibinfo{booktitle}{{\em Software-Intensive Systems and New
  Computing Paradigms - Challenges and Visions}}. \bibinfo{series}{LNCS},
  Vol.~\bibinfo{volume}{5380}. \bibinfo{publisher}{Springer},
  \bibinfo{pages}{101--115}.
\newblock
\showDOI{%
\url{https://doi.org/10.1007/978-3-540-89437-7_6}}


\bibitem[\protect\citeauthoryear{Tan, Vuran, and Goddard}{Tan
  et~al\mbox{.}}{2009}]%
        {TVG09}
\bibfield{author}{\bibinfo{person}{Y. Tan}, \bibinfo{person}{M.~C. Vuran},
  {and} \bibinfo{person}{S. Goddard}.} \bibinfo{year}{2009}\natexlab{}.
\newblock \showarticletitle{Spatio-Temporal Event Model for Cyber-Physical
  Systems}. In \bibinfo{booktitle}{{\em 2009 29th IEEE International Conference
  on Distributed Computing Systems Workshops}}. \bibinfo{publisher}{IEEE},
  \bibinfo{pages}{44--50}.
\newblock
\showISSN{1545-0678}
\showDOI{%
\url{https://doi.org/10.1109/ICDCSW.2009.82}}


\end{thebibliography}

\end{document}